\documentclass[journal,comsoc]{IEEEtran}

\usepackage[T1]{fontenc}
\usepackage{amsmath}
\usepackage[cmintegrals]{newtxmath}
\usepackage{comment}
\usepackage{xspace}
\usepackage{xcolor}
\usepackage{todonotes}
\usepackage{quoting}
\usepackage[font={bf,it,small},labelsep=colon]{caption}
\usepackage{enumitem}
\usepackage{hyperref}
\usepackage{flushend}

\newcommand{\eg}{{e.g.,}\xspace}
\newcommand{\ie}{{i.e.,}\xspace}

\iffalse
\newcommand{\TODO}[1]{\todo[inline,color=brown!40]{TODO: #1}}
\definecolor{aqua}{rgb}{0.0, 1.0, 1.0}
\newcommand{\saurabh}[1]{\vspace{1em} \todo[inline,color=aqua!40]{saurabh: #1}}
\definecolor{amber}{rgb}{1.0, 0.75, 0.0}
\newcommand{\ran}[1]{\vspace{1em} \todo[inline,color=amber!40]{ran: #1}}
\newcommand{\tarek}[1]{\vspace{1em} \todo[inline,color=blue!40]{tarek: #1}}
\newcommand{\ramesh}[1]{\vspace{1em} \todo[inline,color=orange!40]{ramesh: #1}}
\newcommand{\prashant}[1]{\vspace{1em} \todo[inline,color=applegreen!40]{prashant: #1}}
\newcommand{\pradipta}[1]{\vspace{1em} \todo[inline,color=yellow!40]{pradipta: #1}}
\else
\newcommand{\TODO}[1]{}
\newcommand{\saurabh}[1]{}
\newcommand{\ran}[1]{}
\newcommand{\tarek}[1]{}
\newcommand{\ramesh}[1]{}
\newcommand{\prashant}[1]{}
\newcommand{\pradipta}[1]{}

\fi

\newcommand\paraspace{\vspace*{0.5ex}}
\newcommand\parab[1]{\paraspace\noindent\textbf{#1}}
\newcommand\parae[1]{\paraspace\textbf{\textit{#1}}}

\begin{document}
\title{New Frontiers in IoT: Networking, Systems, Reliability, and Security Challenges}

\author{Saurabh Bagchi,
        Tarek F. Abdelzaher,
        Ramesh Govindan,
        Prashant Shenoy, 
        Akanksha Atrey,  
        Pradipta Ghosh,
        and Ran Xu
\thanks{Saurabh Bagchi is with the School of Electrical and Computer Engineering and the Department of Computer Science at Purdue University in West Lafayette, IN, USA 47907, e-mail: sbagchi@purdue.edu.}
\thanks{Tarek F. Abdelzaher is with the Department of Computer Science at University of Illinois at Urbana Champaign in Urbana, IL, USA 61801, e-mail: zaher@cs.uiuc.edu.}
\thanks{Akanksha Atrey is with the College of Information \& Computer Sciences at University of Massachusetts in Amherst, MA, USA 01003, e-mail: aatrey@cs.umass.edu.}
\thanks{Pradipta Ghosh is with the Department of Computer Science at the University of Southern California in Los Angeles, CA, USA 90089, e-mail: pradiptg@usc.edu.}
\thanks{Ramesh Govindan is with the Department of Computer Science at the University of Southern California in Los Angeles, CA, USA 90089, e-mail: ramesh@usc.edu.}
\thanks{Prashant Shenoy is with the College of Information at University of Massachusetts in Amherst, MA, USA 01003, e-mail: shenoy@cs.umass.edu.}
\thanks{Ran Xu is with the School of Electrical and Computer Engineering at Purdue University in West Lafayette, IN, USA 47907, e-mail: xu943@purdue.edu.}
}

\markboth{IEEE IoT Journal}%
{A \MakeLowercase{\textit{et al.}}: Computer Systems and Networking Challenges in IoT}

\maketitle



\begin{abstract}
The field of IoT has blossomed and is positively influencing many application domains. In this paper, we bring out the unique challenges this field poses to research in computer systems and networking. The unique challenges arise from the unique characteristics of IoT systems such as the diversity of application domains where they are used and the increasingly demanding protocols they are being called upon to run (such as, video and LIDAR processing) on constrained resources (on-node and network). We show how these open challenges can benefit from foundations laid in other areas, such as, 5G cellular protocols, ML model reduction, and device-edge-cloud offloading. We then discuss the unique challenges for reliability, security, and privacy posed by IoT systems due to their salient characteristics which include heterogeneity of devices and protocols, dependence on the physical environment, and the close coupling with humans. We again show how the open research challenges benefit from reliability, security, and privacy advancements in other areas. We conclude by providing a vision for a desirable end state for IoT systems. 
\end{abstract}


%
\IEEEpeerreviewmaketitle

\section{Introduction}
\IEEEPARstart{I}{oT} is an interdisciplinary field as evidenced from the breadth of disciplines that contribute techniques to this field. It involves hardware, software, and often humans, with resource constraints on the hardware (cost, complexity, energy sources) and the software (complexity, compute and memory footprint, disconnected mode of operation). 
To focus the discussion, let us lay down a working definition of IoT, and the ways it is different from the allied disciplines of cyber physical systems, networked control systems, and embedded systems. 

\begin{quoting}
The {\em Internet of Things} refers to networked devices that interact with their physical surroundings and communicate over wireless networks in social contexts to offer a human-centric application value.
\end{quoting}

Accordingly, concerns in IoT intersect with cyber-physical systems in that the system may contain embedded components and may include associated control algorithms. However, IoT systems are by definition distributed, putting more emphasis on end-to-end systems challenges, scalability, and network support within the end-to-end application context, as opposed to, say, control systems. Also, IoT systems, by virtue of distribution and scale, are often multipurpose. As such, specific capabilities may be put together dynamically, leading to challenges in composability and integration.  

\parab{Application context.}
IoT application areas fall into three categories:
\begin{enumerate}
    \item Enhance our spaces, in which humans live (e.g., homes and offices).
    \item Empower the devices we use (e.g., appliances, vehicles).
    \item Improve the efficiency of production and delivery systems (e.g., agriculture, the power grid, manufacturing) so as to improve human life and productivity.
\end{enumerate}
An important aspect of these applications is the human in the loop, to generate sensor readings (e.g., crowd-sourcing), to validate control decisions, or to act upon the actuation commands. 

\parab{Challenges and constraints.}
There are some key technical challenges that are salient to the IoT domain. 
There are often constraints on hardware and software that preclude heavy-duty computation (such as, expensive asymmetric cryptographic operations) or significant storage overhead  (such as, a large ensemble of models). There is often a constraint on the wireless networking available to the nodes --- it is often low data rate and there may also be periods of disconnected operation, such as, due to wireless brownouts or interference from multiple devices operating in a public ISM band. There is often a real-time constraint on the tasks, else financial loss or human discomfort may occur, such as, inefficient electricity use in an industrial setting or uncomfortable indoor environment. The heterogeneity of devices and the corresponding wireless protocols they support pose challenging engineering problems. For example, one device may have access to trusted hardware such as ARM TrustZone while the majority of devices may not have such hardware; some may speak only Zigbee as a short-range wireless protocol and not have the long-range cellular or LoRa stack, while other nodes may have the capability for long-range communication. Finally, the human in the loop brings forth challenges for operation (must be simple enough in the parts where human interaction is needed), maintainability (must not require complex or frequent maintenance operations), and safety (must not endanger human users).

In this work, we present the broad open challenges in IoT, from the computer systems and networking aspect and from the reliability, security, and privacy aspect. Within each, we first look at the foundations that we can build on in terms of analytical, architectural, and systems building blocks already available to us. 

\section{Systems and Networking Challenges}

\subsection{Unique challenges}

IoT systems required substantial systems design innovation, primarily because  of: (a) the diversity of sensors and actuators with different wireless technologies they use, (b) the variety of indoor and outdoor locations they are deployed in, (c) the unpredictable conditions under which they are deployed, including unpredictability in availability and quality of network connectivity, (d) the interaction of humans in the loop, and (e) the energy and compute power constraints. 

In recent years, researchers have explored a wide range of challenges related to IoT networks with small, battery-powered sensors. At present, we can concede that we understand that space well (\cite{yick2008wireless,wang2006reprogramming}). The next phase of IoT research will focus on analytics and control using richer sensors that provide various forms of visual information: camera, radar, LiDAR, stereo cameras etc. These ``visual''\footnote{For simplicity, we call these visual sensors, because they can ``see'' the environment, albeit in different ways than humans might in some cases.} sensors provide semantically rich information, but can require significant processing to extract this information. Equally important, with decreasing cost and form factors, sensors like cameras and LiDARs are being deployed densely, even on personal mobile devices.  These will enhance the quality of decisions for most applications discussed above. 
Beyond richer sensors, future IoT systems will include autonomous drones and vehicles that add significant complexity to control and actuation. Constraints imposed by compute and network will be the primary bottlenecks in realizing the full potential of these future IoT systems.

\begin{figure*}[t]
  \centering
  \includegraphics[width=0.8\textwidth]{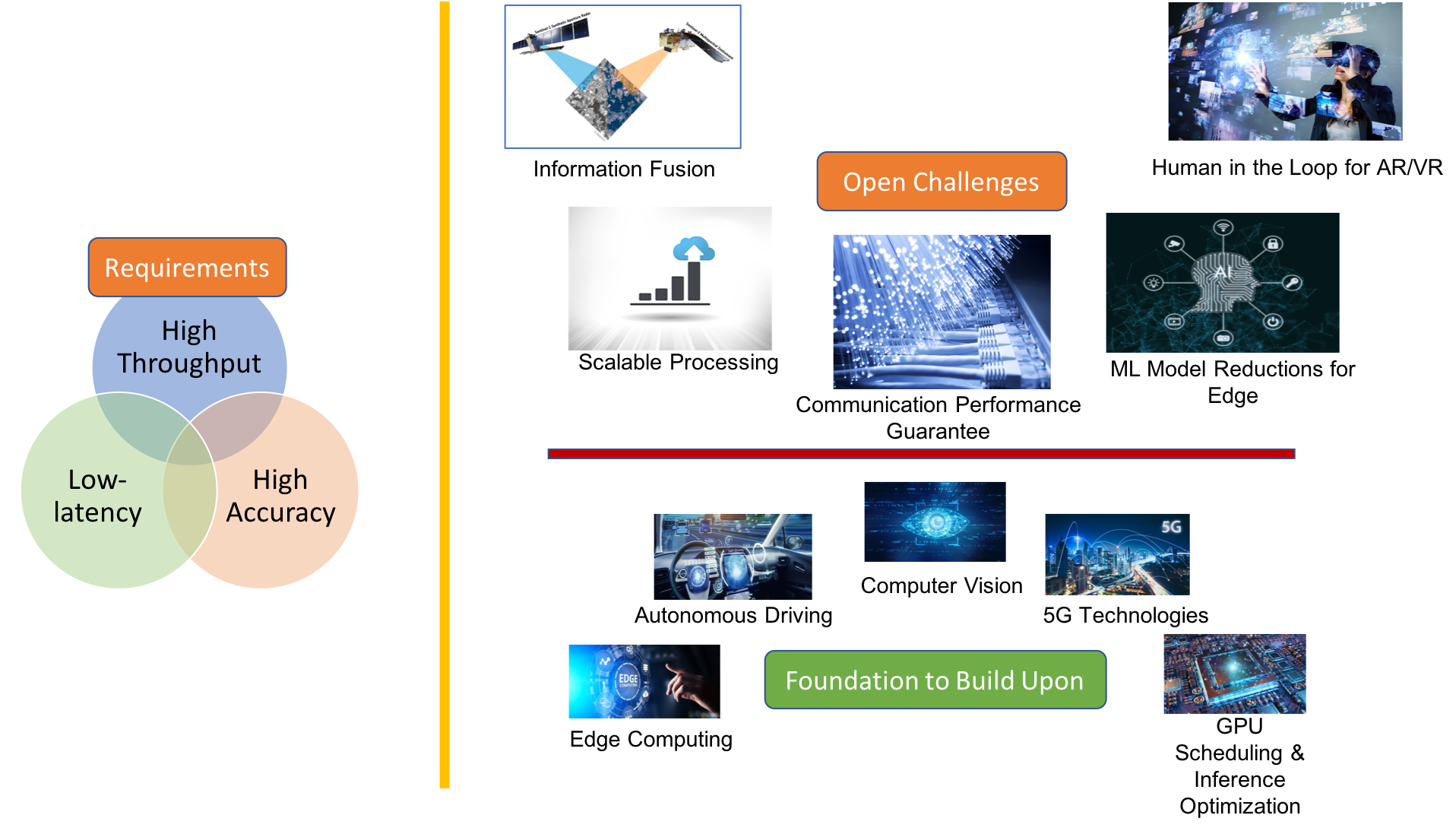}
  \caption{Overview of the novel systems and networking requirements and challenges in IoT systems and foundations from current work that we can build upon.}
  \label{fig:net_sys}
\end{figure*}

\subsection{Performance Requirements}

Before delving into the challenges, we first identify (Figure~\ref{fig:net_sys}) some key requirements for an IoT network consisting of sensors (like cameras and LiDARs), and actuators (like drones and vehicles).

\parab{High throughput/frames.} While a low power temperature sensor generates a few bytes of data every minute, a camera or LiDAR can generate data at several hundred Mbps. Their devices are often connected via wireless interface to a cloud or edge cluster to process the frames. Transmitting raw sensor data may be infeasible given even future wireless standards. 
 
\parab{Low latency.} For actuating a drone or a vehicle, an IoT system will need to support ultra-low end-to-end latencies on the order of a few milliseconds. The two main sources of latency in the IoT control loop are: processing the sensor data, and communication delay. 

\parab{High accuracy.} Especially for visual sensing, accuracy is an essential performance metric. For example, the accuracy of sensing and tracking objects  can affect vehicular or drone control significantly.

Future IoT systems will need to simultaneously satisfy all of these requirements, and visual sensors, together with near real-time control of vehicles and drones, represent extreme points in the space of requirements across all three dimensions.


\subsection{Research challenges.} We now describe new research challenges that arise as a result of the three requirements mentioned above. 


\parab{Information Extraction and Fusion.} Future IoT networks will include a wide range of sensors in terms of sensing frequency and the amount of data they generate. \emph{Rich sensors} like cameras and LiDAR produce significantly more information than a low-end thermostat or an accelerometer. In order to fuse sensors meaningfully, we need to extract only useful and non-redundant information in a timely manner. Combining sensing information from different sensing modalities is an extremely challenging task~\cite{hall1997introduction}. Existing systems fuse 2-3 different types of sensors \eg camera with LiDAR \cite{song2016robust, kato2002obstacle, cornick2016localizing}: but do not generalize to a large number (both in type and count) of heterogeneous rich sensors. The main challenges related to information extraction and fusion in IoT with a large number of \emph{rich sensors} are as follows.

\parae{Data Registration.}  Incorporating data from different scans or sensors to generate a unified view is often known in the computer vision and autonomous vehicles community as data registration~\cite{habib2005photogrammetric}. Data registration deals with properly combining a large number of 3D point clouds obtained from 3D sensors such as stereo cameras and LiDAR where each sensor generates a point cloud with respect to its own frame of reference. 3D point clouds tends to be very large in size (from hundreds of Megabytes to hundreds of Gigabytes) and thus cannot be exchanged between devices. On the other hand, such point clouds tend to provide very fine-grained and extremely dynamic sensing information. Therefore the timeliness and accuracy of data registration is important for any IoT applications that rely on the combined point cloud. 

\emph{Foundations to Build Upon:} Data registration has been studied for many years in the context of autonomous vehicles and robotics maneuvers and path planning~\cite{liu2016linear, bellekens2014survey}. Most existing work relies on the fact that the sensors are co-located or located at close proximity (on the vehicle or robot) with significant overlap in the sensing regions. In the context of IoT, roadside 3D sensors may cover a large area with very small or limited overlap in the sensing area~\cite{zhao2019detection}. For such situations, existing solutions may be inadequate. Recent work has started exploring the problem of data registration in the context of infrastructure (roadside) LiDARs~\cite{yue2019data,zhao2019exploring}. Nonetheless, these solutions do not address data registration for roadside 3D sensors at scale and are limited to only 2-3 sensors at maximum. State-of-the-art methods also lack data registration techniques involving heterogeneous IoT infrastructure sensors such as LiDAR and Stereo Cameras.

\parae{Detection/classification/tracking techniques.} Existing computer vision  detection/classification/tracking techniques for camera and LiDAR processing are typically resource-heavy, limited to a small number of devices, and performed on cloud infrastructure~\cite{dong2009advances}. In an IoT environment significant innovation is required for real-time sensing across multiple devices such as tracking activity across a large number of overlapping and non-overlapping cameras~\cite{liu2019caesar} or LiDARs. 

\emph{Foundations to Build Upon:} Object detection, classification, and tracking are active areas of research in computer vision~\cite{ragland2014survey,brunetti2018computer}. These tasks are usually performed by training a deep neural network (DNN) with a large dataset catered towards a particular detection, classification, or tracking task. For illustration, let us consider the task of real-time human activity detection in live camera streams. There exist a large number of monolithic DNN models~\cite{ulutan2018actor, shou2018online, gowsikhaa2012suspicious} that extract features from video streams and predict actions of every human appearing in the video. To detect interactions, a class of methods analyze the moving trajectories of objects near a person to predict the interaction between the person and the object~\cite{mettes2017spatial, amor2016action} while another class of methods opt to train separate DNNs to detect group behavior such as ``walk in group'', ``stand
in queue'' (\cite{bagautdinov2017social, amer2014hirf}). While these methods perform well for a small number of detection/classification/tracking tasks (limited by the availability of relevant datasets), they cannot be tailored for any tasks outside the vocabulary and thus are not generalizable for future IoT operations. 
Even with existing DNN solutions for specific activities, human intervention is required for analyzing complex activities  that involve specific activities being detected by the DNN, e.g, ``two men chatting, then exchanging a document, then walking in a group.'' Moreover, the processing time of the video steams increases exponentially on shared computation resources as the number of tracked objects increases~\cite{liu2019caesar}. Additional challenges appear as we scale such detections across multiple heterogeneous devices. To perform tracking across multiple video streams, we need proper synchronization of the images frames, timely processing of image frames, and re-identification of object/humans across multiple cameras. Existing work has explored the single-camera action detection~\cite{ulutan2018actor, shou2018online, zhao2017temporal}, tracking of people across multiple overlapping cameras~\cite{xu2017cross,nithin2017globality,solera2016tracking} and non-overlapping cameras~\cite{ristani2018features, tesfaye2017multi,chen2017equalized}. However, due to complexity of the problem, very few researchers have looked into  a real-time generic tracking and detection across multiple cameras which is required for future IoT systems~\cite{liu2019caesar}. Similar observations can be made for almost all kinds of detection/classification/tracking state-of-the-arts. Looking forward, significant innovation and development is required towards detection/classification/tracking technique that are generalizable for a large vocabulary of tasks dealing with a large number of heterogeneous imaging devices. One way to achieve this is to take advantage of the existing DNN solutions and combine them in a semantically meaningful, systematic way as shown in \cite{liu2019caesar}.

%
\parae{Compute and Communication Constraints.} Processing the data stream from one camera/LiDAR is a challenging task itself. For an IoT network with multiple cameras/LiDARs, the processing time and resource requirements are very high~\cite{zhang2017live}. This calls for innovations at the algorithmic level to process a large number of sensor data streams with accuracy and processing time similar to processing of a single sensor data stream. This has to be accomplished on platforms that are not as resource rich as server-class platforms and where the isolation guarantees among applications are weaker. 
In addition, transmitting multiple sensory data streams to a cloud or edge requires lot more additional communication bandwidth than supported by typical shared wireless medium such as WiFi~\cite{wang2018bandwidth}. 

\emph{Foundations to Build Upon:} 
Scalable processing of video and LiDAR data stream is a cutting edge field of research. Both types of data require resource-heavy CNN/DNN to extract the embedded rich information. Future IoT networks will include a large number of heavy sensors such as cameras, LiDAR for smart sensing. While some application require processing of a combined data (via data registration, explained above), others require concurrent and separate processing of individual data streams on shared compute resources. Often, based on the task query, one might need to run multiple different DNN on same video stream. Most of the future IoT applications will rely on a chain of DNNs running on edge clusters.
Researchers have looked into this problem in the context of live video streaming from multiple cameras~\cite{zhang2017live,hung2018videoedge}. The state-of-the-art live-stream processing systems operate knobs for different performance settings (frame rate and resolution) to maximize the shared server utilization and maintain a minimum quality of service for all task queries. 

While downgrading quality of frames is a potential scalable option, it often results in lowering the accuracy of the DNN/CNN. A more recent class of approaches employ GPU multi-tenancy scheduling on TensorFlow Serving platforms~\cite{tfserving} to improve GPU sharing and utilization~\cite{hu2018olympian} on shared edge cloud. Some state-of-the-art techniques also save GPU cycles by caching intermediate results~\cite{crankshaw2017clipper, lee2018pretzel}, lazily activating DNN~\cite{liu2019caesar}, and  batching the input for higher per-image processing speed on GPU~\cite{crankshaw2017clipper}. However, all these solutions work well for a small range of applications for lower frame rate and a small number of concurrent streams (<10) and concurrent queries, and cannot scale to large numbers of concurrent streams.  This is relevant because we anticipate that future IoT systems will include a large number of concurrent streams, multiple edge clusters, and  a large number of DNNs (or chains of DNN) per image stream. To this, we need to remove any redundancy present in the input, DNN, or GPU schedule. Identifying a set of sensing-objective specific key-frames (instead of processing every frame) is essential and is a promising field of research in this context.

Beyond video, recent work has started exploring the design of vehicular IoT systems that use depth perception sensors. For example, AVR~\cite{Qiu18d} has explored a combination of several techniques, including dynamic object extraction and adaptive transmission of stereo camera point clouds to enable extended vehicular vision. Similarly, CarMap~\cite{CarMap} efficiently uploads updates to high-definition maps over a cellular network, using a combination of techniques to produce a lean map representation that does not sacrifice positioning accuracy.

\parae{Accuracy vs performance tradeoff.} To support communication and processing of multiple data streams from \emph{rich sensors} (such as cameras and LiDARs) with a limited shared resources (bandwidth, GPU etc), researchers often adopt techniques to drop data frames (randomly or selectively) by keeping a set of key-frames~\cite{wang2018bandwidth}. Such approaches tend to achieve the performance requirement in terms of throughput (goodput) and runtime at a cost of reduced accuracy. However, to achieve certain throughput one needs to achieve accuracy above a certain threshold. Thus, the tradeoff between accuracy and performance requires careful analysis and consideration for designing systems and algorithm for future IoT. Specifically, simple application-agnostic techniques like frame dropping may not suffice to achieve good accuracy; often, application-specific techniques that leverage problem structure to extract only information essential to the problem~\cite{CarMap,Qiu18d} can provide orders of magnitude performance improvement while minimally impacting accuracy.


\parab{The role of edge/cloud offload, and device computation.}
Edge computing~\cite{shi2016edge} will play a central role for future IoT networks. Often the on-board processing power of a camera/LiDAR device is unable to run necessary processing pipelines (deep learning models) 
to extract the embedded rich information.  This calls for offloading the computation either to a cloud with large processing power at the cost of larger unpredictable delays or to a nearby edge device, or a cluster of edge devices, with enough processing power and with lower, more predictable delays. This raises a series of questions: Which option should we choose, edge or cloud, or a hybrid? What data to share with the edge/cloud? How to minimize the end-to-end latency of the processing pipeline while reducing the communication overhead? The future also presupposes the possibility of having multiple heterogeneous edge devices, some of which are unmanaged while the rest are managed (by commercial organizations). Unmanaged edge implies that such devices are voluntarily contributed by the public and are unpredictably available. 

Of particular interest in this context, is the introduction of machine intelligence into the IoT edge/cloud architecture~\cite{yao2019eugene}. IoT will push the boundaries of federated learning motivated by the fact that each individual device may be too resource constrained and by privacy requirements in IoT settings. Neural networks offer a great portable representation, much like a language virtual machine (e.g., Java and Python), that makes it possible to distribute inference algorithms across edge and cloud machines, and control the amount of communication among them. 
Services might (i) generate deep neural network models (from client-supplied training data), (ii) help with (automatic) labeling of data sets, and (iii) perform model reduction (if needed for caching on the edge device). Generated models might be executed as appropriate on the server or client. This vision poses several challenges.

\parae{Model Reduction for IoT Devices:}
Modern machine intelligence algorithms are heavy-weight. To run on a low-end IoT device, solutions are needed to reduce the computational and memory needs of machine inference. Recent work shows that model reductions of orders of magnitude are possible~\cite{yao2017deepiot,liu2018demand}. For example, a device can cache a reduced model that identifies a number of most frequent commands, leaving the more general (but rarely encountered) identification tasks to the cloud. Models can also be customized to the specific hardware. For example, rather than minimizing computational cost, a model that fully utilizes an available GPU will give a better quality/consumption trade-off~\cite{yao2018fastdeepiot}.
Alternatively, the end device may choose to offload the inference to a server.
Communication between the resource-constrained IoT end-device and an edge/cloud-processing server will need to be compressed~\cite{dey2019offloaded}. Auto-encoder-like solutions allow asymmetric encoding/decoding where the IoT-device-side encoder (that compresses the data onto a lower-dimensional manifold) is lightweight, whereas the decoder (running on an edge-server) is more involved. Order-of-magnitude reduction in communication was shown using compressive offloading~\cite{yao2019stardust}.
On the server, since improvements in result accuracy diminish with increased depth of the neural network, efficiency considerations suggest that once a desired quality is achieved, the service should refrain from executing additional layers. A scheduler may determine how many stages to execute to avoid diminishing returns. 

\parae{Data Prioritization:}
A commonly overlooked challenge in IoT-centric machine inference contexts is one of data prioritization. When a human driver observes a scene, they instinctively prioritize regions of higher criticality in the scene (such as a child on the side of the road who might run across at any instant) over regions of lower criticality (such as buildings in the background, fire hydrants, trees, etc). No such prioritization is done in current machine learning software. Rather, some of the heaviest computational operations are performed on all bits of an image in every frame without prioritization. A novel stack is needed that is aware of importance of different regions in an image. Some examples were proposed in recent literature.  

\parab{Communication Requirements.}  IoT networks heavily rely on wireless communication for inter-device communications. State-of-the-art wireless communication technology needs to accommodate for the high throughput and low-delay requirements of future IoT networks involving cameras and LiDARs.
Camera or LiDAR data streaming via state-of-the-art wireless communication standards experience many challenges affecting the performance such as unnecessary re-transmission, bandwidth fluctuation due to dynamic channel quality, lack of dedicated channel access due to contention-based MAC protocol, and heterogeneous devices sharing that same medium~\cite{hsiao2011h}. 
The situation is likely to become more adverse by the incorporation of Augmented Reality (AR) and Virtual Reality (VR) devices in future IoT networks. A single VR device requires hundreds of Megabytes to couple Gigabytes of dedicated bandwidth for a reasonable user experience~\cite{bastug2017toward}. 
While data compression and coding techniques~\cite{wark2007real} can help to reduce the bandwidth requirements, current wireless networking technologies still fall short of fulfilling the bandwidth and performance guarantee requirements.
Moreover, in a wireless network with a large number of heterogeneous devices (cameras, LiDARs, etc.) and actuators (drones, autonomous cars), the network needs to have provision for prioritizing certain types of traffic such as control traffic as well as maintain fairness and performance guarantees among multiple data streams. 

\emph{Foundations to Build Upon:}
The fifth-generation network (5G) is the obvious core wireless technology for the future IoT network as it will allow for higher datarate (up to tens of GBps) which is orders of magnitude higher than current wireless technologies~\cite{qi2016quantifying}. To this, researchers have started to explore 5G based communication architectures for future IoT~\cite{ojanpera2018use}. While 5G offers significantly more bandwidth and data rates, we still require technologies to offer precise control of traffic and performance in the shared wireless medium. To meet the performance demands by facilitating the flexible allocation of resources, future IoT networks must make use of recent networks virtualization concepts such as software-defined network (SDN)~\cite{bizanis2016sdn,tayyaba2017software}, network functions virtualization (NFV)~\cite{mijumbi2016management}, and network slicing~\cite{alliance2016description}. 

In addition, modifications are required in streaming protocols (e.g., video) to reduce unnecessary information and save bandwidth~\cite{pakha2018reinventing,zhang2017live,hung2018videoedge}. 
Conventional streaming protocols (such as RTMP~\cite{parmar2012adobe} for video) and encoding standards (such as H.265~\cite{hsiao2011h} for video) are tailored towards maximizing user quality of experience (QoS). 
Such protocols tend to optimize the frame rate and resolutions to avoid unnecessary interruptions and delays. In future IoT networks, the majority of streaming will be tied to analytics where the objective is to maximize the inference accuracy and the performance objectives are different from normal live streaming. For example, in a video analytics application,  frame resolution beyond a threshold has a negligible effect on the DNN/CNN based object detection pipelines and often only a small cropped portion of the frames are used~\cite{liu2019caesar, pakha2018reinventing}. In addition, sequential frames in a video stream might not have any additional information and can be dropped to save bandwidth without incurring any performance deterioration. Thus, video streaming protocols for future IoT analytics have many parameters to tune for such as frame selection --- area cropping (and transmitting only the cropped area), resolution of the image, and compression that are relatively unexplored in existing video streaming protocols. Similar scope of research lies in other types of streaming applications such as LiDAR data streaming, and audio streaming. 

\parab{Humans in the Loop.} A key distinguishing feature of IoT systems is the human in the loop. Humans consume the output of IoT systems, but may also provide inputs inputs to add reliability and context to IoT systems~\cite{nunes2015survey}. While such intervention by humans in IoT systems has its advantages, modeling and analysis of these IoT systems require modeling of human behavior. This is particularly challenging due to the complex physiological, psychological, and behavioral aspects of human beings. Apart from the human modeling aspect, there also exist several system design challenges such as minimizing human input and coping up with occasional unpredictability and unreliability of human inputs. The set of challenges is even broader in the context of AR/VR applications for future IoT networks. Consider a battlefield IoT setting where relevant roadside camera/LiDARs streams are live-fed to the VR headset of ground troops. The quality of streaming and the switching between different infrastructure sensor feeds heavily rely on the soldier input. The main challenge there is to associate the correct infrastructure sensors by comparing the live stream from the infrastructure sensors and the live-stream from head-mounted camera of the soldier. Such reliable association requires a combination of DNN-based pipelines and inputs from the soldiers and is currently an open area of research.

\subsection{Path forward}

We have to solve the above research challenges through coordinated optimization of the compute and communication that is spread out among a diverse set of resources. This will be helped by open architecture for IoT that standardizes sensing and actuation and distributed computation. In all our solutions, we have to design for humans as first order entities interacting with the rest of the system elements. 

\section{Reliability Challenges}

With IoT systems being deployed in critical application areas, including in those where human safety is at stake, reliability is an important and hitherto rather neglected aspect of IoT systems. We discuss the unique requirements and challenges plus the foundations from current work that we can build upon. 

\begin{figure*}
\centering
\includegraphics[width=0.80\textwidth,clip]{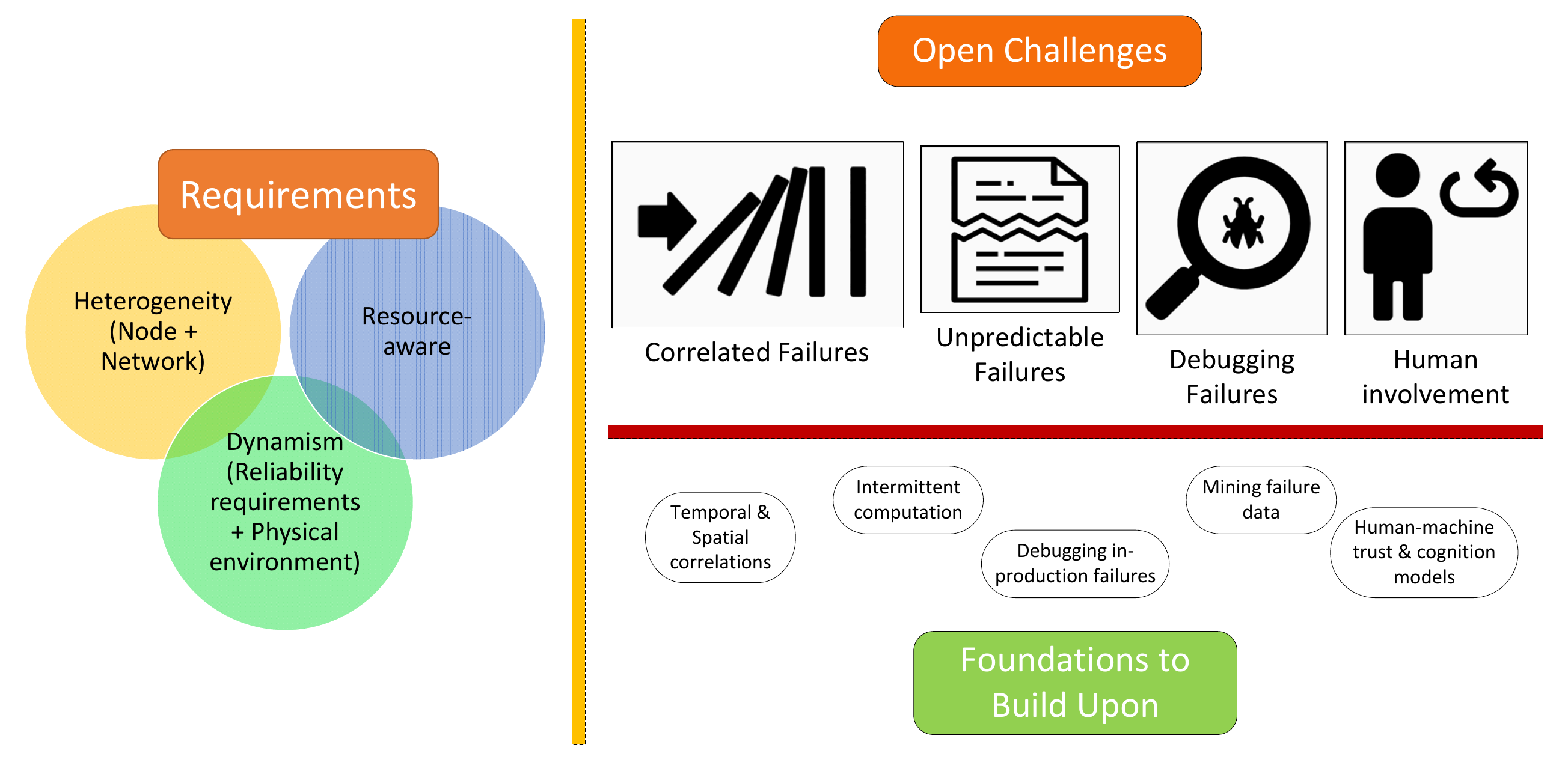}
\caption{Overview of the novel reliability requirements and challenges in IoT systems and foundations from current work that we can build upon}
\label{fig:reliability-overview}
\vspace{-6mm}
\end{figure*}

\subsection{Unique Challenges}
As IoT systems have become more than playthings and are deployed in applications with moderate to high criticality requirements, they require reliable architecture, operation, and application development. The reliability must address errors in the hardware, the software, interactions with the physical environment, and interactions with the human users. One development is that the systems generate large volumes of data, often at high rates, which put new pressure on the reliability mechanism. The data can be of mixed criticality (i.e., some of it is critical and if not properly handled, lead to user-visible failures, while the rest of it is not) and therefore heterogeneous reliability processing is called for. As introduced earlier, heterogeneity is a first order feature of our target systems. This heterogeneity also applies to the reliability area, both in design and operation. For example, some devices have software developed through rigorous software development practices and in programming languages that are safe by design, while some others may have agile software development in unsafe programming languages. Further, due to the runtime instantiation, different devices have different capacity for tolerating errors---some component may be capable of masking errors, while others propagate the errors.  
Finally, the real-time aspect of the operation implies that reliability measures cannot perturb the timing too much. While there is a mature design and development of reliability for hard real-time systems, our target systems pose new challenges because they are developed much faster (e.g., with little to no formal validation) and they operate in more diverse and uncertain environments. 
Related to the issue of reliability is the notion of predictable behavior from the system, despite the presence of multiple unpredictable factors---in the IoT platform, in interactions among the platforms, and in interactions between the system and the human users. This is important since the IoT system often has human-in-the-loop or human-on-the-loop (the former means human {\em has} to be involved in the chain of decision making while the latter makes that optional). Humans have varying degrees of aversion to uncertainty and this underlines the need for this aspect of system operation for IoT systems. 

\subsection{Requirements}
It is necessary for the reliability protocols to be diverse, in keeping with the heterogeneity of the runtimes where they will execute and heterogeneity of the applications that they are meant to protect. The reliability protocols should be adaptive, to the current state of resources on the device (\eg a resource-intensive but critical task may start up on the device), the current reliability requirement (\eg the current data stream being gathered, processed, and communicated to the back-end may be highly critical for some downstream application), and the current state of the physical environment (\eg a physically hazardous environment may cause correlated failures of multiple devices in spatial proximity). 

\subsection{Research challenges}

There are four broad themes in the salient research challenges that face reliability of IoT systems. 

\parae{Handling correlated failures.} This involves dealing with failures that are correlated in space and time. Spatial correlation occurs due to the fact that multiple devices may face similar physical or cyber environments, such as, wireless congestion or high temperature fluctuation. Temporal correlation occurs due to some physical phenomenon spreading with time and affecting devices serially, such as, high moisture content causing device failures, or the coordinated movement of a large mass of people causing excessive number of concurrent events.  
    
\parae{Handling unpredictable failures.} A large fraction of failures are unpredictable in any system. This effect is magnified in IoT systems due to several factors. First, the energy resources get drained in an unpredictable manner, say due to environmental conditions for rechargeable solar battery, or unanticipated load leading to high communication activity. Second, an IoT system does not have much headroom when it is deployed, \ie there is not much safety factor that is built into their deployment. So even mildly aberrant conditions, such as small spikes in load, can cause the system to go into a tailspin leading to failures. Third, there does not exist as good modeling of the failure modes of these systems, as for server-class systems. 
    
\parae{Debugging failures}. It is important to enable automated debugging of failures in IoT systems, with the stress on automation due to the fact that the systems are made of a large number of heterogeneous devices, which would stress human cognition for debugging. Automated debugging is challenging because not all execution data can be logged at the devices and not all the logged data can be communicated to a back-end for debugging. Further, distributed debugging is often needed, bringing together traces from multiple devices.
    
\parae{Human considerations.} This reliability challenge arises due to the human-in-the-loop (or on-the-loop) in many of these IoT systems. This means different things in different applications and even different deployments for the same application. For example, some human users may be highly reluctant to endure false alarm rates, while some human users may be loathe to look at alarms on small form factor displays on devices. A typical human-centric form of unreliability arises when human users are distracted while interacting with the systems. The issue of maintainability is inextricably related to this theme, whereby it is important that these systems can be maintained (upgraded, re-flashed, reconfigured, etc.) with little to no human intervention, and hardly any expert intervention. 
    

\subsection{Foundations to Build Upon}
For each of the above themes, there is sparse to moderate amount of work that is ongoing. We survey some of the most promising works in each. 

First, on the theme of {\em handling correlated failures}, researchers have developed a rich set of methods to detect faulty sensors and architecture to improve the reliability of IoT systems. 
The work in~\cite{bronevetsky2012automatic} uses the insight that with correlated failures, elementary detectors will flag many events almost coincident in time. The authors show that a single-level ML classifier underperforms for many realistic system-level faults, while having a two-stage detection (clustering events at the first stage) improves the detection and false positive rates considerably. 
%
To handle space-correlated failures, 
%
%
%
Bychkovskiy {\em et al.}~\cite{bychkovskiy2003collaborative} present a two-phase post-deployment calibration technique for large-scale sensors. The key idea is to use the temporal correlation of signals in the co-located sensors and maximize the consistency among the groups of sensor nodes.
Balzano {\em et al.}~\cite{balzano2007blind} whether proposes blind calibration approach for sensor networks from weakly correlated sensor readings.
%
%
On the other hand, focusing on the network connectivity, Neumayer {\em et al.}~\cite{neumayer2010network} develop tools to model and analyze geographically correlated network failures.
%
As for temporally correlated failures, Sharma {\em et al.}~\cite{sharma2010sensor} propose time series analysis-based methods to detect faulty sensors. 
%
Jeffery {\em et al.}~\cite{jeffery2006declarative} present a framework, called Extensible Sensor stream Processing (ESP), to clean the both time and space correlated sensor data in the pervasive applications.
%
Apart from the space and time correlated failure, Szewczyk {\em et al.}~\cite{szewczyk2004lessons} find that failure of temperature sensors is highly correlated with the failure of the humidity sensors in their lessons from a sensor network expedition.
%
%
Researchers from data mining community also provide valuable analytic models for such co-related sensor data. Dong {\em et al.}~\cite{dong2009integrating} considers the dependence between data sources in truth discovery where the conflicting information may come from a large number of sources.
%
Although lots of models have been proposed to clean sensor data, calibrate sensor reading and detect sensor faults, we have not seen much work that uses the recent machine learning approaches for failure detection with correlations.

Second, on the theme of {\em handling unpredictable failures}, a line of solutions have been applied to energy-harvesting IoT devices where failure can happen unpredictably due to energy drain. Some work in this space~\cite{maeng2017alpaca, van2016intermittent} inserts checkpoints in the code to save state that the application can recover from. Some advanced work~\cite{maeng2018adaptive} does the checkpointing based on available energy. %
Lightweight approaches are presented in some recent studies. Karimi {\em et al.}~\cite{karimienergy} 
present a new energy scheduling scheme to execute periodic real-time tasks on the intermittently-powered embedded devices. Maeng {\em et al.}~\cite{maeng2020adaptive} also present the adaptive low-overhead scheduling for intermittent execution.
However, the open questions center around how to handle a {\em larger set} of unpredictable failures in a manner that respects the currently available resources (available storage, energy, etc.). 

Third, on the theme of {\em debugging failures}, most compelling works rely on collecting runtime information and deducing anomalous behavior automatically by mining patterns in the information. Unfortunately there is a lack of workable solutions for debugging in-production failures. One promising direction is record and replay, whereby execution traces are recorded on the devices when the system is operational and the traces are somehow brought back to a backend for replaying and debugging. Within the realm of record 
and replay, our prior work Tardis ~\cite{tancreti2015tardis} was the first software-only record and replay system for 
embedded devices. However, it is only applicable to a single node and does not consider execution 
on the commonly used microcontrollers (e.g., those which run multi-threaded OSes and applications). 
Our work Aveksha~\cite{tancreti2011aveksha} uses extra hardware to record traces from the JTAG port without interfering 
with the node’s execution, but cannot record complete control flows. This and other approaches like 
Minerva~\cite{sommer2013minerva} and FlashBox~\cite{choudhuri2009flashbox} that use hardware modifications cannot be deployed to 
COTS IoT systems. Some software-only efforts, such as TinyTracer~\cite{sundaram2011demo} and Prius~\cite{sundaram2012prius}, selectively 
record some events (only control flow for TinyTracer) and therefore cannot enable replay-based 
debugging. The open questions center around how to provide high fidelity system-level replay, 
i.e., replay that is able to reproduce both control flow at an instruction level and
the state of memory at any point in time for any software module executing on the node. On the broader theme of uncovering patterns in the traces, there needs to be learning algorithms that can learn such patterns from observations in the field. Such solutions will take the place of current, fragile rule-based approaches. 

Fourth, on the theme of {\em human considerations in reliability}, researchers have developed approaches to create more reliable systems and identify failures when error occurs. 
%
As human operators get involved in the control loop of IoT sensor networks, Gross {\em et al.}~\cite{gross2017supervisory} use tandem human-machine cognition approach to mitigate and avoid cognitive overload situations where false alarms and ambiguity may overwhelm humans. 
Humans can also affect the connection between smart things. Thus, Guo {\em et al.}~\cite{guo2013opportunistic} use opportunistic IoT models to enable information forwarding and dissemination within the opportunistic IoT communities which are formed based on the movement of humans. 
In the context of Social IoT applications and services, Truong {\em et al.}~\cite{truong2016reputation} develop a Trust Service Platform with a trust model incorporating both reputation properties and knowledge-based property so that multiple entities can trust each other.
On the other hand, to identify potential causes for human failures, Cranor~\cite{cranor2008framework} proposes a framework for reasoning about the human in the loop in the secure system.
%
However, the open questions include how to use a unified model to study the human factors in the reliability of IoT systems, considering the large variety of humans, perhaps with machine learning models.

\section{Security and Privacy Challenges}

With the large volume of data generated by sensors and large number of heterogeneous IoT devices, some embedded in our private secure physical spaces, security and privacy pose new challenges. We discuss each of these  individually below.

\subsection{Security}

\begin{figure*}
\centering
\includegraphics[width=0.80\textwidth,clip]{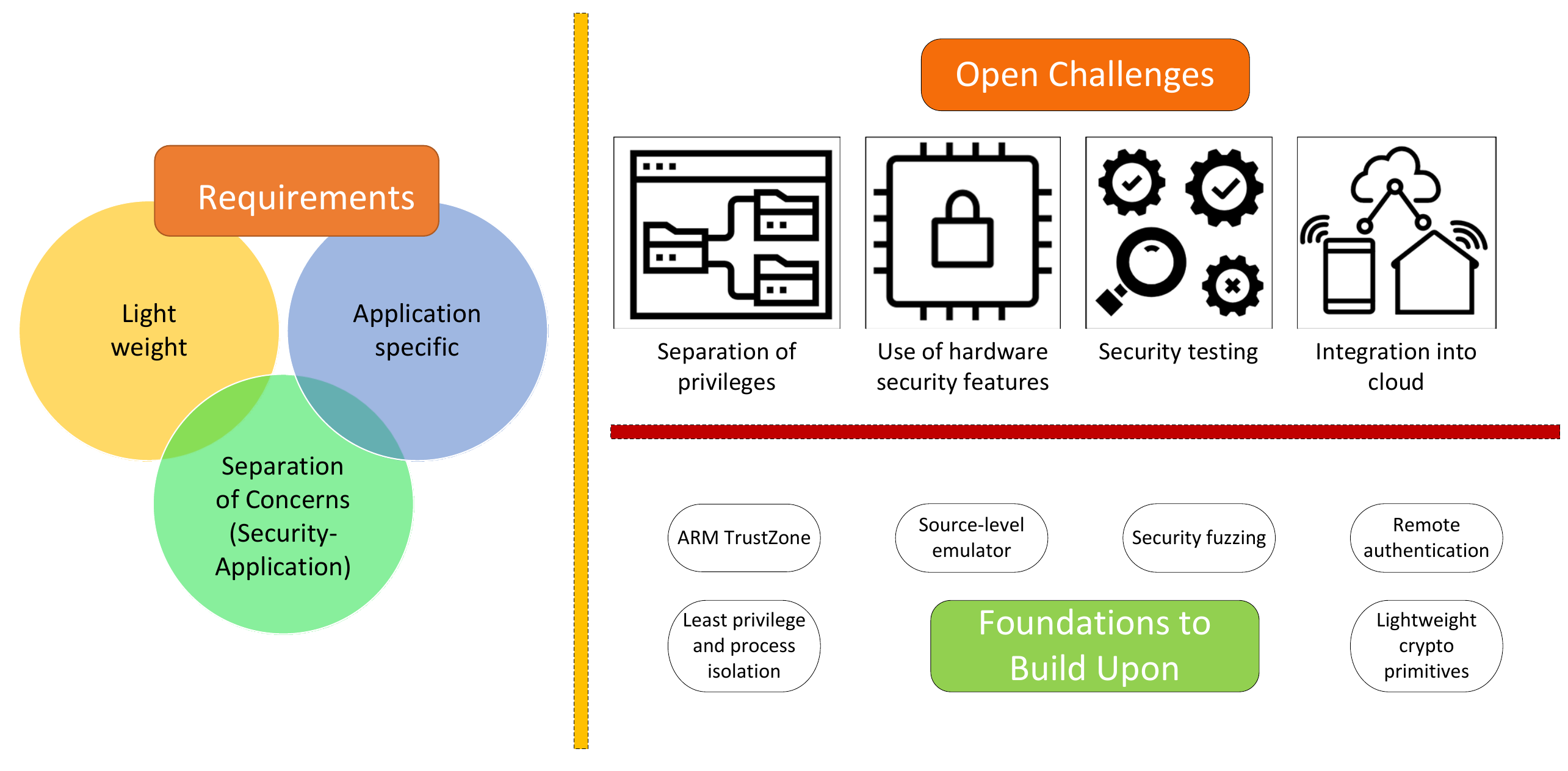}
\caption{Overview of the novel security requirements and challenges in IoT systems and foundations from current work that we can build upon}
\label{fig:security-overview}
\vspace{-6mm}
\end{figure*}

\subsubsection{Unique Challenges}
The proliferation of increasingly connected devices
has led to new levels of connectivity
and automation in IoT systems.  The connectivity has great potential
to improve our lives, however, it exposes such systems to network-based
attacks on an unprecedented scale.  Attacks against IoT devices
have already unleashed massive Denial of Service attacks~\cite{Krebs2016DDoS},
given hackers access to streaming video feeds deep inside the periphery of a corporate IT network~\cite{iot-video-attack-2018}, taken control of
autonomous vehicles~\cite{car-hacking-2019}, and facilitated robbing hotel
rooms~\cite{HotelLockHack}. Currently, these devices are deployed
with no security mitigations against a wide variety of attacks that are commonly
expected in server-class systems. For example, Data Execution Prevention (DEP) is a fundamental and widely adopted security primitive in server-class systems, whereby all writeable memory pages are marked as non-executable---this is often also referred to as the \texttt{W$\oplus$X} defense~\cite{DEP}. But this relies on special hardware in the CPU (AMD “NX” bit (no-execute), Intel “XD” bit (executed disable)), which is often not present in IoT CPUs. 
As another example, consider Address Space Layout Randomization (ASLR) whereby the memory address layout is randomized from one instance of an application to another instance of the same application~\cite{bhatkar2003address}. The goal is that ASLR prevents attackers from using the
same exploit code effectively against all instantiations of the
program containing the same flaw. However, this relies on a certain degree of randomness such that a brute-force attack will take a long time to succeed and such randomness relies on a large memory space~\cite{shacham2004effectiveness}, which is often not available on our target systems. 
When security defenses {\em are} present in IoT systems, mitigations are often implemented in an ad hoc manner, relying on the developer to make good security decisions. Therefore, such defenses are easily bypassable, \eg by writing a single flag value to disable all memory protections~\cite{clements2017protecting}.
We posit that {\bf as IoT devices become ubiquitous, security must become a first class principle}.

\subsubsection{Requirements}
Security in our target systems must be able to fit inside the available hardware and software and must not perturb the timing properties significantly, neither increasing significantly the mean execution time or even the variance in it. It must provide clear separation of concerns between the application development and the security development so that the application developer is not called upon to make subtle security design decisions. This is challenging particularly due to the fact that security configuration here is often {\em application-specific}. For example, an IO
register on one system may unlock a lock while on a different
system, it may control an LED used for debugging. Clearly the
former is a security-sensitive operation while the latter is not.
To balance the two factors, such application-specific requirements should be supported
in a manner that does {\em not} require the developer to make
intrusive changes within her application code. Finally, and perhaps most importantly, the security techniques and their instantiations must be easily portable across different systems. Such portability should apply say within the same vendor's family of products \eg within ARM M-class microcontrollers, despite the presence of a different and heterogeneous set of peripherals from one system to the next. 

\subsubsection{Research challenges}
There are four broad themes in the research challenges that face the security of IoT systems.

\begin{enumerate}[leftmargin=0em,itemindent=1.5em]
    \item {\em Separation of privileges}. IoT devices no
longer focus on a dedicated task but increasingly run
multiple independent or loosely related tasks. For example,
a single SoC often implements both Bluetooth and
WiFi, where neither Bluetooth nor WiFi needs to access
the code and data of the other. However, without isolation,
a single bug compromises the entire SoC and possibly
the entire system (one demonstration was taking over Android smartphones through compromising Broadcom's Wi-Fi SoC~\cite{artenstein2017broadpwn}). It is important to bring in the notion of least
privileges or process isolation to the IoT systems. The first notion refers to the need to grant each software component the minimum privilege needed to complete its functionality, while the second refers to the need to protect the control and the data flow from an unprivileged component affecting a privileged component. The research efforts in this theme need to achieve these while respecting the requirements laid out above. This is a challenge because the overwhelming majority of existing IoT software is written with the assumption that any software module can access any other software module or hardware block, \ie there is no notion of separation of privileges. It is complex to first identify the different functional software modules (software in this domain is often deeply tangled) and then it is difficult to figure out what is the right set of privileges to assign to each module. A paramount concern is not to break existing functionality and thus avoid significant porting costs. 
    
    \item {\em Effective use of hardware security features}. 
    While high-end trusted hardware solutions like Intel SGX are typically considered not feasible for large-scale IoT deployments, there are widely used hardware-based trusted execution environments through features like ARM TrustZone. At a more universal level, most micro-controllers come equipped with a
peripheral called the Memory Protection Unit (MPU) that can enforce read, write, and execute permissions on regions of the physical memory. TrustZone is also being pushed down into lower end devices, such as, ARM Cortex-M microcontroller series. The challenge is to use such hardware features efficiently and securely. From an efficiency standpoint, consider that the number of MPU registers is limited, \eg the latest generation, ARM Cortex v8-M processors, have 13 MPU registers. This means that the security protection granularity has to be appropriately defined at any point in the execution to fit within these many registers. 
For the TrustZone-based solutions, typically applications have to be rewritten using the particular API, which imposes a burden, an insurmountable one at times, for adoption.  For the security consideration, it is important for the solution to be such that it cannot be bypassed by an out-of-band mechanism that simply disables the use of the security hardware. For example, for MPU protection, it can be disabled simply by writing a 0 to the lowest bit of the \texttt{MPU\_CTRL} register, which is at a fixed (and therefore, known) memory address. For ARM TrustZone, security challenges arise due to the desire to share the device among multiple applications. It is important to guarantee that the isolation among the applications is preserved even when each makes use of the TrustZone. 

    \item {\em Security testing}. 
    Simply verifying the security guarantees of these IoT systems is often a challenge in the face of blackbox software packages. Thus, standard mechanisms for verifying security properties like symbolic verification cannot be brought to bear on IoT software. Today, developers create and test IoT firmware almost entirely on physical testbeds, typically consisting of development versions of the target devices. However, modern software engineering practices that benefit from scale, such as test-driven development, continuous integration, or fuzzing, are challenging or impractical due to this hardware dependency~\cite{muench2018you}. In addition, embedded hardware provides limited introspection capabilities, including extremely limited numbers of breakpoints and watchpoints, significantly restricting the ability to perform dynamic analysis on firmware. Manufacturing best-practices dictate stripping out or disabling debugging ports (e.g., JTAG), meaning that many off-the-shelf devices remain entirely opaque. Even if the firmware can be obtained through other means, dynamic analysis remains challenging due to the complex environmental dependencies of the code, such as, dependency on the specific version of a garden variety peripheral like an Ethernet card.
    
    \item {\em Secure integration of IoT into cloud services}. 
    As there is an increasing drive to integrate IoT devices into cloud services, it is essential from a security standpoint to be able to validate the security properties of the devices, at startup as well as periodically, say before doing any critical operation involving these devices. For this, there are three classes of techniques that need to be developed. First, is {\bf remote authentication} whereby any IoT device being brought online is properly authenticated. This should stay away from using sources of information that are low entropy (or equivalently easily guessed), such as the MAC address (MAC addresses of devices are often allocated based on the vendor and the high order bits are publicly known). A second class of techniques is {\bf remote attestation (RA)}, which involves verification of current internal state (\ie RAM and/or flash) of an untrusted remote hardware platform (an IoT device in our context) by a trusted entity (say, a service running on the cloud on behalf of an end user). RA will allow for devices to be compromised, but a remote verification can uncover the presence of malware or other effects of such compromise. This has to be done in a way that balances the resource usage on the device and the security guarantees (either formal or empirical) that the scheme can provide. The third class and broader of techniques relates to the use of {\bf crypto primitives} on these resource-constrained platforms. It is important that the crypto primitives fit within the resource budget of the device, chiefly memory and energy, but provide rigorously quantified security guarantees. Only then can higher level security protocols that integrate these devices with the cloud be built up. Too often in the past have there been cases of insecure design or implementation of crypto primitives for IoT-class of devices, \eg WEP for wireless transmissions (insecure design)~\cite{cam2003security} and car keyless entry (insecure use of crypto keys)~\cite{garcia2016lock}. This is a particularly pressing research challenge in this domain because of the ease of eavesdropping on communication, due to the omnidirectional wireless communication channel, and the difficulty of upgrading software (including crypto software) once devices are deployed in the field. 
    
\end{enumerate}

\subsubsection{Foundations to Build Upon}

On the theme of {\bf Separation of privileges}, FreeRTOS-MPU provides privilege separation between user tasks and kernel task~\cite{freertos-mpu}. However, there is a significant barrier to usability in that the separation has to be carefully and manually programmed in by the application developer. Some other approaches~\cite{clements2017protecting, clements2018aces, kim2018securing} use static and
dynamic analysis to enforce separation of privileges between different compartments
of IoT software allowing a system owner to enforce the principle of least privileges, which is a bedrock of security. Such approaches break the single application into smaller
compartments and enforce data integrity and control-flow integrity between compartments.
Specific open questions are how far can the separation be done automatically, what is the relative role of static and dynamic techniques, what is the interplay between performance overhead and security in any compartmentalization, and how does a given design overlay on the available hardware features of the device. It is probably unarguable that we have far to go for the compartmentalization to reach the level of sophistication we have on server-class software and systems and work on all three fronts --- programming frameworks, tools for using such frameworks, and runtime environments --- will help us get there. 

On the theme of {\bf Effective use of hardware root of trust}, techniques such as EPOXY~\cite{clements2017protecting} and ACES~\cite{clements2018aces} make it impossible for the hardware root of trust to be configured (including bypassed) from any but small amount of privileged code. For ARM TrustZone compatibility, some solutions present a sophisticated runtime environment that shields the applications from the TrustZone API thus ensuring that legacy applications can be supported~\cite{guan2017trustshadow}. For the secure sharing of the TrustZone, some solutions have been developed that provide secure virtualization and isolation among multiple applications~\cite{hua2017vtz}. The broad open question relates to how much application modification is tolerable---if the modification can be templated, then that process can be automated. A second question relates to the efficiency loss due to the intervening layer that tries  to support legacy applications. Also, since the runtime has not been scrutinized to the extent that the TrustZone TEE has been, are there security bugs lurking there? 

On the theme of {\bf Security testing}, there are several promising directions that attempt to address one or a few of the challenges mentioned above. The promising line of work here is emulation using source code where available and binary blobs for other parts~\cite{zaddach2014avatar, chen2016towards, clements2020halucinator}. The source code is executed either on the actual hardware or a source-level emulator such as QEMU and the binary blobs are analyzed through mature binary analysis tools like IDA, Ghidra, or LibMatch and then re-hosted on a standard emulator (thus alleviating the pain point that the actual esoteric version of a peripheral may not be available during testing). Then all standard software testing techniques can be brought to test the execution on the emulator. These works differ in the layer of the binary at which they do the analysis (high-level libraries vs lower-level), the fidelity of the analysis (do they give up when they encounter a binary blob without symbol table or can they perform approximate analysis), and the dependence on hardware-in-the-loop (what kind of hardware do they need to execute). The broad open questions that we need to tackle are how much manual effort is needed in testing IoT software---is manual understanding of the black-box binary blobs needed or can that be replaced with simple input-output behaviors, do fuzzing and symbolic execution engines need to be equipped with domain-specific constraints. Another broad question is does the fact that testing perturbs the timing of the application change the kinds of bugs that it exposes.

\begin{figure*}[ht!]
\centering
\includegraphics[width=0.80\textwidth,clip]{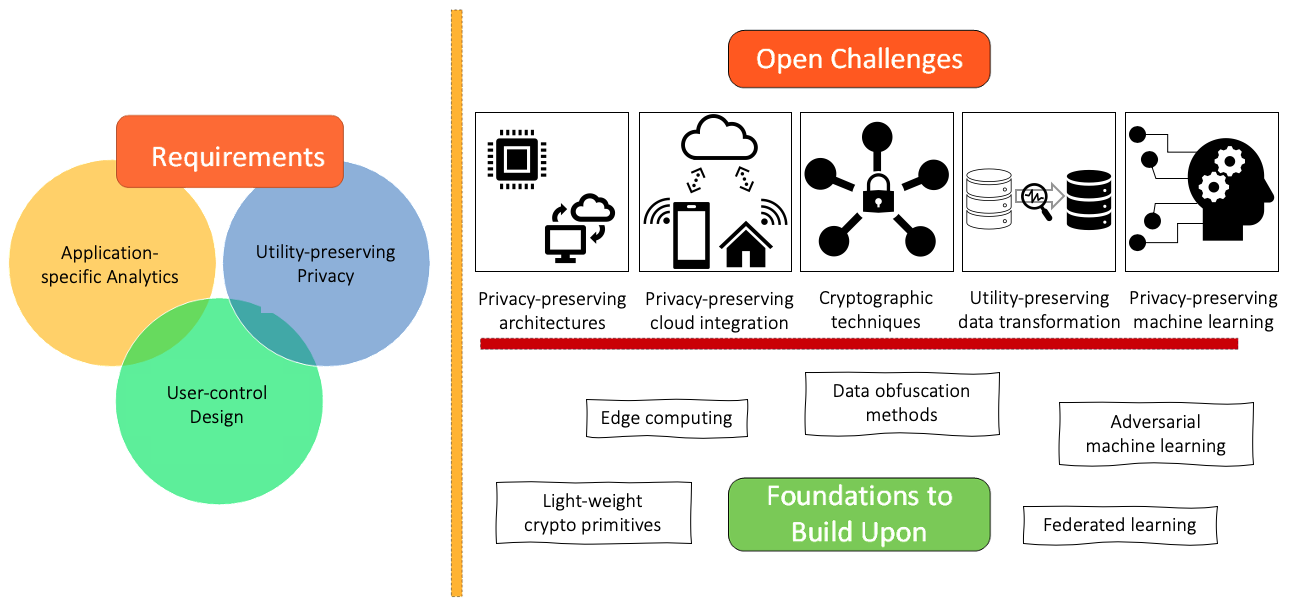}
\caption{Overview of the novel privacy requirements and challenges in IoT systems and foundations from current work that we can build upon.}
\label{fig:privacy-overview}
\vspace{-6mm}
\end{figure*}

On the theme of {\bf Secure integration of IoT devices into cloud-based systems},~\cite{chen2019your} brings to light questionable security practices with 10 IoT vendors for remote binding of IoT devices to cloud services, in the designs of authentication and authorization, including inappropriate use of device IDs, weak device authentication, and weak cloud-side access control. It brings forth a fundamental problem with building authentication from hard-coded device attributes like device ID. Such attributes may be leaked through device ownership transfer, including device reuse, reselling, stealing, and so forth. One possible approach is to "refresh" these sources of information through remote reprogramming, either periodically or based on critical events (such as, change in location). Such reprogramming can erase old state or increase the entropy of the variable being relied on for authentication. However, reprogramming has to be done keeping in mind the network constraints of latency and bandwidth and several solutions exist in that space~\cite{panta2011efficient}.  Several authentication and authorization platforms for IoT have been built~\cite{fernandes2016security, tian2017smartauth}, which differ in the usability, the granularity of the control, and the kinds of devices they can be run on. Several schemes for remote attestation have been built~\cite{eldefrawy2012smart, ibrahim2017seed, nunes2019vrased}, which differ in what kinds of devices they are targeted at (very low-end TI MSP430 class or higher end ARM R-class), are they hardware-based (such as using TrustZone), software-based (\ie based on timing properties), or hybrid, and how formally they have been modeled and verified. The broad open question is how best to combine secure protocols for bringing devices online and remotely managing them including detecting compromise or verifying their integrity. This has to be done while ensuring that any crypto primitive being relied on has enough entropy to be able to withstand cryptanalysis attacks for the required duration of time (the time duration itself may be very application dependent).

\subsection{Privacy}

Privacy is another important topic since IoT devices are embedded in our physical spaces including privacy-sensitive locations.

\subsubsection{Unique Challenges}
As IoT devices become more pervasive, they have begun to collect data about our environment, our homes, our health and many other aspects of our lives. This data may contain sensitive or private information that needs to be safeguarded from the user's perspective. As an example, consider smart voice assistants that listen for spoken commands or smart cameras that continuously view our home environment. Safeguarding the privacy of IoT data raises new challenges that go beyond traditional data privacy challenges.

The issue has become important due to the proliferation of consumer IoT products that range from smart outlets, smart door locks, thermostats, cameras, fitness bands, voice assistants, object trackers and many more. Unfortunately, many of these products are designed to provide convenience to users (eg. remote operation) but often pay little attention to user privacy.

As has been noted earlier, the current generation of consumer IoT products use a cloud-based architecture where data generated by the device is sent to the cloud for processing~\cite{chen2018private}. The cloud service can then run any analytics on this data to derive insights for the user. In effect, the user no longer has full control over this data and it is possible for the cloud provider or third parties to mine this data for private information.

\subsubsection{Requirements}
Privacy researchers have proposed data obfuscation as a possible approach to ensure privacy of user data in the cloud \cite{mclaughlin2011protecting, yang2012minimizing}. Data obfuscation involves transforming the data by adding noise to it prior to transmitting it to the cloud. While obfuscation methods can ensure better privacy, they are a blunt instrument. The transformed data reveals nothing after obfuscation and is no longer useful for performing cloud analytics. That is, obfuscation removes all information embedded in the data, both private and non-private.
Hence, privacy-preserving techniques for IoT data need to carefully consider what type of private and non-private information is present in the data and determine how to mask private information without hampering the ability to perform useful analytics on the data. Further, allowing users greater control over their privacy is a key design requirement.

\subsubsection{Research Challenges}
\begin{enumerate}[leftmargin=0em,itemindent=1.5em]
    \item {\em Privacy-preserving architectures.} While current IoT devices rely on a cloud-based architecture, researchers have begun to study new architectures that have better privacy properties \cite{grossmann2019applicability, grossmann2020cloudless}.
    For instance, "cloudless" architectures that process all IoT data on the device itself or an edge device located on customer premises are emerging. These new architectures are becoming feasible due to the rapid hardware advantages that have resulted in specialized chips (eg. Apple's Neural Engine~\cite{apple2017neural} and Intel's Movidius VPU~\cite{intelmovidius}) that allow sophisticated computation to be performed on low-end hardware. For instance, such chips have allowed some security cameras to perform face recognition on-device and without sending video data to the cloud.
    A key advantage of such architectures is that the data is retained by the user and stays on user premises where it is processed locally. Thus, third-parties do not have access to the data and can no longer mine it for sensitive information.
    
    \item {\em Privacy-preserving integration into cloud services.} The previous section described challenges in secure integration of IoT into cloud services. Security and privacy are related but distinct challenges. Even with secure cloud integration, IoT services do not necessarily provide privacy. This is because cloud analytics on secure IoT data can still leak private user information.
    
    Consequently, privacy preserving techniques are needed in addition to security techniques for cloud-based IoT services. Some works have attempted to develop novel cloud-based architectures and integration techniques that preserve IoT privacy \cite{funke2015end, jayaraman2017privacy}. The main challenge is to design techniques that thwart side-channel attacks. Side channel attacks essentially mine or infer orthogonal information from the original purpose for which the data was collected. For instance, electricity usage data recorded by electric meters are known to reveal occupancy information based on periods of higher usage, a type of side channel attack~\cite{koo2017rl}. The problem is especially challenging since it is a priori unclear what kind of other information may be hidden within the data gathered for a specific purpose. Conventional techniques such as differential privacy often do not apply since we are concerned with masking private information from a single data stream.
    
    \item {\em Cryptographic techniques for IoT privacy.} Analogous to crypto-based security methods, researchers have developed cryptographic primitives for ensuring privacy when transmitting data to cloud services \cite{danezis2011towards}. One such approach leverages zero knowledge cryptography (ZKC), where the IoT device sends a cryptographic proof to the cloud server rather than the raw data~\cite{molina2010private}. Such a proof, known as a zero knowledge proof, allows the server to verify that the result was derived from valid data.
    However, each ZK proof is based on a specific query and general methods that allow for a broad set of analytic queries to be performed with ZKC in the IoT context remains an open challenge.
    
    \item {\em Utility-preserving data transformation.} An alternate approach to ZKC is to employ intelligent data transformation on the data prior to transmitting it to the cloud. Unlike obfuscation-based methods that leave no useful information in the data, utility-preserving privacy transformation seek to mask any type of private information in the data while leaving other non-private information intact. Doing so allows conventional analytics in the cloud to be performed like before, but prevents side channel attacks that try to extract private information from the data. Such utility-preserving privacy transformation are more challenging than data obfuscation since they need to selectively mask only information considered to be private. Existing methods such as differential privacy are useful on multi-user data \cite{dwork2008differential}. However, since analyzing single-user streams is more prominent in the IoT context, novel utility-preserving techniques are required.
    
    Utility-preserving privacy also raises an interesting trade-off between utility and privacy. The more information that is masked in the data, the less useful it becomes (obfuscation can be considered to be an extreme case that masks all information). Thus, it is important to consider user preferences when designing such systems and let the user decide what information to suppress and what to reveal to a cloud service. For instance, Zheng {\em et al.}~\cite{zheng2018user} use semi-structured interviews with smart home owners to understand their reasons for purchasing IoT devices, perceptions of smart home privacy risks, and actions taken to protect their device and data privacy. This is as much of a HCI challenge as a technical one since explaining privacy implications to users to choose the appropriate preferences is a non-trivial issue.
    
    \item {\em Privacy-preserving machine learning.} There has been a growing use of machine learning (ML) in the IoT domain. From using a Long Short Term Memory (LSTM) model on smart watch data to detect medical conditions such as diabetes and high blood pressure \cite{ballinger2018deepheart} to detecting distributed denial of service (DDoS) attacks in IoT traffic using random forests \cite{doshi2018machine}, researchers are employing advanced ML methods to improve the usability of IoT devices. 
    
    However, the popularity of ML with IoT data raises a fresh set of privacy concerns. Adversaries with IoT data can employ ML methods to infer private information that may be implicit in the data. 
    For instance, prior work has demonstrated that it is feasible to disaggregate energy usage from smart energy meters into individual components, popularly known as Non-Intrusive Load Monitoring (NILM), using a Factorial Hidden Markov Model (FHMM) \cite{kolter2011redd, batra2014nilmtk}. This type of disaggregation directly reveals daily activity patterns of users. Privacy attacks are also possible on trained ML models. Two such popular attacks are membership inference, where an adversary attempts to infer whether a user was part of the training data, and model inversion, where an adversary attempts to infer sensitive features in the training data via model output. Recent work has shown that membership inference attacks can be conducted on aggregate location data from smart services \cite{pyrgelis2017knock}. Model inversion attacks pose a higher concern from an IoT perspective where distributed ML models can be reverse engineered to gain sensitive local information.
    
    These issues raise an overriding question: how can we use ML to improve usability of IoT devices while preserving privacy? Designing ML models that are privacy-preserving and are robust to model-based attacks in an IoT context is thus a pressing open area of research.
\end{enumerate}

\subsubsection{Foundations to Build Upon}
Edge computing will be a major foundation for future privacy-preserving architectures and privacy-preserving cloud integration. As computation and storage on distributed edge clusters advances, edge-based architectures will gain prominence and cloud integration will involve more aggregate and/or processed data. Lightweight crypto primitives will serve as a foundation for resource-bounded, privacy-preserving cryptography methods for IoT.

Though blunt, data obfuscation methods provide a reasonable foundation towards utility-preserving privacy for cloud-based architectures. Building on data obfuscation methods to mask only private features is an open research direction. Some work has demonstrated success in masking private data in the energy domain \cite{chen2013nonintrusive, chen2014combined, koo2017rl}. Recent work has employed Metropolis-Hasting statistical sampling to transform data from smart energy meters to suppress private user information while retaining non-private information~\cite{phuthipong2020repel}. Such methods can be used towards developing more general utility-preserving techniques.
Recent work in federated learning, a ML technique that allows decentralized training on edge devices, has demonstrated possibilities for privacy-preserving ML in the IoT domain \cite{jiang2019differentially, zhao2019mobile, imteaj2019distributed, sattler2019robust, lu2019blockchain, li2019federated, yu2020learning}. Federated learning-based methods will gain prominence in analyzing IoT data as edge computing and distributed learning advances. Another promising approach is the combination of traditional differential privacy methods with ML to protect aggregate user data~\cite{liu2018epic, xiong2018enhancing, xu2018distilling}.

\subsection{Path forward}

Reliability and security have quickly become important for IoT systems as they are being used in applications where human health or safety or large financial gains/losses are at stake. Privacy is a unique challenge here as IoT systems are embedded in our physical spaces and interact with us through multiple modalities (speech, vision, touch, etc.). We want to make the research and development of these as first order concerns. Their design and development must be enabled in a way that the application developer does not also have to become an expert in them, but rather clean usable interfaces allow understanding and configuration of these building blocks. In terms of these building blocks, they have to be designed and developed in way that they are usable, fit within the resource constraints of the devices and the network, and meet the application-specific goals to different and configurable levels of fidelity. 
\section{Discussion and Take-aways}


IoT systems serve applications that fall under three broad categories: applications that enhance our physical spaces (homes and offices), applications that empower the devices we use (e.g., appliances and vehicles), and applications that enhance the efficiency of production and delivery systems (e.g., food production, manufacturing, and energy delivery). These applications are demanding IoT systems that are {\em simultaneously} high performing, secure, and reliable. IoT systems are distributed, putting more emphasis on end-to-end systems challenges, scalability, and network support, as opposed to, say, control systems or embedded systems. Also, IoT systems, by virtue of distribution and scale, are often multipurpose and heterogeneous and involve humans in the loop. Each of these lead to unique challenges in systems and networking. 

There are a host of promising solutions that are being developed, and with a growing pace of innovations. These include edge-cloud offload and on-device computation, model reduction and efficient model inferencing, 5G and networking software innovations (like Network Function Virtualization), and human-in-the-loop design. These need to be targeted to the unique requirements, focused to work within the constraints, and provide the appropriate interfaces for the human users. 

While developing these solutions, reliability, security, and privacy have to be built into these solutions as first-class primitives. Here also, there are a host of domain-specific challenges. In the area of reliability, promising solutions arise from techniques for dealing with temporally and spatially correlated failures, intermittent computation, debugging in-production failures, discerning failure patterns by mining failure data, and models for human-machine interaction and human cognition. These have to be focused to handle correlated and unpredictable failures (including those due to the close coupling of the devices with the physical environment), debugging large-scale production failures, and reliability bottlenecks due to humans in the loop. In the area of security, several existing solution approaches can be leveraged and many of these are under active development now. These include efficient use of hardware root of trust, enforcement of least privilege and process isolation, remote authentication, lightweight crypto primitives, and security fuzzing for uncovering vulnerabilities. These need to be further developed to reach the goals --- allow IoT devices to be securely integrated into the cloud infrastructure, allow separation of concerns in software development between security and application functionality, and enforce security containment boundaries. These have to be achieved even though hardware features that we rely on in the server world, such as, memory management units, are often not present here. 

Privacy is a particularly important concern since IoT devices are embedded in our physical spaces including in privacy-sensitive locations. Data obfuscation, on-premises processing of IoT data, and privacy-preserving ML are important building blocks for reaching the goals of privacy. These raise an overriding question: How can we use ML to improve usability of IoT devices while preserving privacy? Designing ML models that are privacy-preserving and are robust to model-based attacks in an IoT context is thus a pressing open area of research.

In summary, this is an exciting time to be working in IoT, in its systems, network, reliability, security, or privacy areas. We see a slew of energizing technical challenges and a mounting set of compelling solutions, with many more to come in the near future. 


\ifCLASSOPTIONcaptionsoff
  \newpage
\fi

{\small
\bibliographystyle{plain}
\bibliography{references,sbagchi,pradipta, aatrey}

\begin{thebibliography}{100}

\bibitem{intelmovidius}
Intel vision accelerator design with intel movidius vision processing unit
  (vpu).
\newblock
  \url{https://software.intel.com/en-us/iot/hardware/vision-accelerator-movidius-vpu#specifications}.

\bibitem{tfserving}
{{T}ensorFlow {S}erving}.
\newblock \url{https://github.com/tensorflow/serving}.

\bibitem{apple2017neural}
Apple’s ‘neural engine’ infuses the iphone with ai smarts.
\newblock
  \url{www.wired.com/story/apples-neural-engine-infuses-the-iphone-with-ai-smarts/},
  September 2017.

\bibitem{CarMap}
Fawad Ahmad, Hang Qiu, Ray Eells, Fan Bai, and Ramesh Govindan.
\newblock Carmap-fast 3d feature map updates for automobiles.
\newblock In {\em 17th USENIX Symposium on Networked Systems Design and
  Implementation (NSDI 20)}, Santa Clara, CA, 2020. USENIX Association.

\bibitem{alliance2016description}
NGMN Alliance.
\newblock Description of network slicing concept.
\newblock
  \url{https://www.ngmn.org/publications/description-of-network-slicing-concept.html},
  2016.

\bibitem{amer2014hirf}
Mohamed~Rabie Amer, Peng Lei, and Sinisa Todorovic.
\newblock {H}irf: {H}ierarchical {R}andom {F}ield for {C}ollective {A}ctivity
  {R}ecognition in {V}ideos.
\newblock In {\em European Conference on Computer Vision}, pages 572--585,
  2014.

\bibitem{amor2016action}
Boulbaba~Ben Amor, Jingyong Su, and Anuj Srivastava.
\newblock {A}ction {R}ecognition using {R}ate-invariant {A}nalysis of
  {S}keletal {S}hape {T}rajectories.
\newblock {\em IEEE transactions on pattern analysis and machine intelligence},
  38(1):1--13, 2016.

\bibitem{artenstein2017broadpwn}
Nitay Artenstein.
\newblock Broadpwn: Remotely compromising android and ios via a bug in
  broadcom’s wi-fi chipsets.
\newblock {\em Black Hat USA}, 2017.

\bibitem{bagautdinov2017social}
Timur Bagautdinov, Alexandre Alahi, Fran{\c{c}}ois Fleuret, Pascal Fua, and
  Silvio Savarese.
\newblock {S}ocial {S}cene {U}nderstanding: {E}nd-to-end {M}ulti-person
  {A}ction {L}ocalization and {C}ollective {A}ctivity {R}ecognition.
\newblock In {\em Proceedings of the IEEE Conference on Computer Vision and
  Pattern Recognition}, pages 4315--4324, 2017.

\bibitem{ballinger2018deepheart}
Brandon Ballinger, Johnson Hsieh, Avesh Singh, Nimit Sohoni, Jack Wang,
  Geoffrey~H Tison, Gregory~M Marcus, Jose~M Sanchez, Carol Maguire, Jeffrey~E
  Olgin, et~al.
\newblock Deepheart: semi-supervised sequence learning for cardiovascular risk
  prediction.
\newblock In {\em Thirty-Second AAAI Conference on Artificial Intelligence},
  2018.

\bibitem{balzano2007blind}
Laura Balzano and Robert Nowak.
\newblock Blind calibration of sensor networks.
\newblock In {\em Proceedings of the 6th international conference on
  Information processing in sensor networks}, pages 79--88, 2007.

\bibitem{bastug2017toward}
Ejder Bastug, Mehdi Bennis, Muriel M{\'e}dard, and M{\'e}rouane Debbah.
\newblock Toward interconnected virtual reality: Opportunities, challenges, and
  enablers.
\newblock {\em IEEE Communications Magazine}, 55(6):110--117, 2017.

\bibitem{batra2014nilmtk}
Nipun Batra, Jack Kelly, Oliver Parson, Haimonti Dutta, William Knottenbelt,
  Alex Rogers, Amarjeet Singh, and Mani Srivastava.
\newblock {NILMTK}: an open source toolkit for non-intrusive load monitoring.
\newblock In {\em Proceedings of the 5th International Conference on Future
  Energy Systems}, 2014.

\bibitem{bellekens2014survey}
Ben Bellekens, Vincent Spruyt, Rafael Berkvens, and Maarten Weyn.
\newblock A survey of rigid 3d pointcloud registration algorithms.
\newblock In {\em AMBIENT 2014: the Fourth International Conference on Ambient
  Computing, Applications, Services and Technologies, August 24-28, 2014, Rome,
  Italy}, pages 8--13, 2014.

\bibitem{bhatkar2003address}
Sandeep Bhatkar, Daniel~C DuVarney, and R~Sekar.
\newblock Address obfuscation: An efficient approach to combat a broad range of
  memory error exploits.
\newblock In {\em USENIX Security Symposium}, pages 291--301, 2003.

\bibitem{bizanis2016sdn}
Nikos Bizanis and Fernando~A Kuipers.
\newblock Sdn and virtualization solutions for the internet of things: A
  survey.
\newblock {\em IEEE Access}, 4:5591--5606, 2016.

\bibitem{phuthipong2020repel}
Phuthipong Bovornkeeratiroj, Srinivasan Iyengar, Stephen Lee, David Irwin, and
  Prashant Shenoy.
\newblock {RepEL}: A utility-preserving privacy system for {IoT}-based energy
  meters.
\newblock In {\em IEEE/ACM International Conference on Internet of Things
  Design and Implementation (IoTDI)}, 2020.

\bibitem{bronevetsky2012automatic}
Greg Bronevetsky, Ignacio Laguna, Bronis~R de~Supinski, and Saurabh Bagchi.
\newblock Automatic fault characterization via abnormality-enhanced
  classification.
\newblock In {\em IEEE/IFIP International Conference on Dependable Systems and
  Networks (DSN 2012)}, pages 1--12. IEEE, 2012.

\bibitem{brunetti2018computer}
Antonio Brunetti, Domenico Buongiorno, Gianpaolo~Francesco Trotta, and
  Vitoantonio Bevilacqua.
\newblock Computer vision and deep learning techniques for pedestrian detection
  and tracking: A survey.
\newblock {\em Neurocomputing}, 300:17--33, 2018.

\bibitem{bychkovskiy2003collaborative}
Vladimir Bychkovskiy, Seapahn Megerian, Deborah Estrin, and Miodrag Potkonjak.
\newblock A collaborative approach to in-place sensor calibration.
\newblock In {\em Information processing in sensor networks}, pages 301--316.
  Springer, 2003.

\bibitem{cam2003security}
Nancy Cam-Winget, Russ Housley, David Wagner, and Jesse Walker.
\newblock Security flaws in 802.11 data link protocols.
\newblock {\em Communications of the ACM}, 46(5):35--39, 2003.

\bibitem{chen2013nonintrusive}
D.~Chen, S.~Barker, A.~Subbaswamy, D.~Irwin, and P.~Shenoy.
\newblock Non-{I}ntrusive {O}ccupancy {M}onitoring using {S}mart {M}eters.
\newblock In {\em BuildSys}, 2013.

\bibitem{chen2014combined}
D.~Chen, D.~Irwin, P.~Shenoy, and J.~Albrecht.
\newblock Combined {H}eat and {P}rivacy: Preventing {O}ccupancy {D}etection
  from {S}mart {M}eters.
\newblock In {\em PerCom}, March 2014.

\bibitem{chen2016towards}
Daming~D Chen, Manuel Egele, Maverick Woo, and David Brumley.
\newblock Towards automated dynamic analysis for linux-based embedded firmware.
\newblock In {\em Network and Distributed System Security (NDSS) Symposium},
  volume~16, pages 1--16, 2016.

\bibitem{chen2018private}
Dong Chen, Phuthipong Bovornkeeratiroj, David Irwin, and Prashant Shenoy.
\newblock Private memoirs of iot devices: Safeguarding user privacy in the iot
  era.
\newblock In {\em International Conference on Distributed Computing Systems
  (ICDCS)}, 2018.

\bibitem{chen2019your}
Jiongyi Chen, Chaoshun Zuo, Wenrui Diao, Shuaike Dong, Qingchuan Zhao, Menghan
  Sun, Zhiqiang Lin, Yinqian Zhang, and Kehuan Zhang.
\newblock Your iots are (not) mine: On the remote binding between iot devices
  and users.
\newblock In {\em 2019 49th Annual IEEE/IFIP International Conference on
  Dependable Systems and Networks (DSN)}, pages 222--233. IEEE, 2019.

\bibitem{chen2017equalized}
Weihua Chen, Lijun Cao, Xiaotang Chen, and Kaiqi Huang.
\newblock {{A}}n {{E}}qualized {{G}}lobal {{G}}raph {{M}}odel-based
  {{A}}pproach for {{M}}ulticamera {{O}}bject {{T}}racking.
\newblock {\em IEEE Transactions on Circuits and Systems for Video Technology},
  27(11):2367--2381, 2017.

\bibitem{choudhuri2009flashbox}
Siddharth Choudhuri and Tony Givargis.
\newblock Flashbox: a system for logging non-deterministic events in deployed
  embedded systems.
\newblock In {\em Proceedings of the 2009 ACM symposium on Applied Computing},
  pages 1676--1682, 2009.

\bibitem{clements2018aces}
Abraham~A Clements, Naif~Saleh Almakhdhub, Saurabh Bagchi, and Mathias Payer.
\newblock {ACES: Automatic Compartments for Embedded Systems}.
\newblock In {\em 27th USENIX Security Symposium (USENIX Sec)}, pages 65--82,
  2018.

\bibitem{clements2017protecting}
Abraham~A Clements, Naif~Saleh Almakhdhub, Khaled~S Saab, Prashast Srivastava,
  Jinkyu Koo, Saurabh Bagchi, and Mathias Payer.
\newblock Protecting bare-metal embedded systems with privilege overlays.
\newblock In {\em 2017 IEEE Symposium on Security and Privacy (SP)}, pages
  289--303. IEEE, 2017.

\bibitem{clements2020halucinator}
Abraham~A Clements, Eric Gustafson, Tobias Scharnowski, Paul Grosen, David
  Fritz, Christopher Kruegel, Giovanni Vigna, Saurabh Bagchi, and Mathias
  Payer.
\newblock {HALucinator: Firmware Re-hosting Through Abstraction Layer
  Emulation}.
\newblock In {\em 29th USENIX Security Symposium (USENIX Sec)}, pages 1--18,
  2020.

\bibitem{cornick2016localizing}
Matthew Cornick, Jeffrey Koechling, Byron Stanley, and Beijia Zhang.
\newblock Localizing ground penetrating radar: A step toward robust autonomous
  ground vehicle localization.
\newblock {\em Journal of field robotics}, 33(1):82--102, 2016.

\bibitem{DEP}
Microsoft Corporation.
\newblock Data execution prevention.
\newblock 2018.

\bibitem{crankshaw2017clipper}
Daniel Crankshaw, Xin Wang, Guilio Zhou, Michael~J Franklin, Joseph~E Gonzalez,
  and Ion Stoica.
\newblock {C}lipper: {A} {L}ow-latency {O}nline {P}rediction {S}erving
  {S}ystem.
\newblock In {\em 14th USENIX Symposium on Networked Systems Design and
  Implementation (NSDI 17)}, pages 613--627, 2017.

\bibitem{cranor2008framework}
Lorrie~F Cranor.
\newblock A framework for reasoning about the human in the loop.
\newblock 2008.

\bibitem{danezis2011towards}
George Danezis and Benjamin Livshits.
\newblock Towards ensuring client-side computational integrity.
\newblock In {\em ACM Workshop on Cloud Computing Security}, 2011.

\bibitem{dey2019offloaded}
Swarnava Dey, Jayeeta Mondal, and Arijit Mukherjee.
\newblock Offloaded execution of deep learning inference at edge: Challenges
  and insights.
\newblock In {\em 2019 IEEE International Conference on Pervasive Computing and
  Communications Workshops (PerCom Workshops)}, pages 855--861. IEEE, 2019.

\bibitem{dong2009advances}
Jiang Dong, Dafang Zhuang, Yaohuan Huang, and Jingying Fu.
\newblock Advances in multi-sensor data fusion: Algorithms and applications.
\newblock {\em Sensors}, 9(10):7771--7784, 2009.

\bibitem{dong2009integrating}
Xin~Luna Dong, Laure Berti-Equille, and Divesh Srivastava.
\newblock Integrating conflicting data: the role of source dependence.
\newblock {\em Proceedings of the VLDB Endowment}, 2(1):550--561, 2009.

\bibitem{doshi2018machine}
Rohan Doshi, Noah Apthorpe, and Nick Feamster.
\newblock Machine learning {DDoS} detection for consumer internet of things
  devices.
\newblock In {\em IEEE Security and Privacy Workshops (SPW)}, 2018.

\bibitem{dwork2008differential}
Cynthia Dwork.
\newblock Differential privacy: A survey of results.
\newblock In {\em International conference on theory and applications of models
  of computation}, 2008.

\bibitem{eldefrawy2012smart}
Karim Eldefrawy, Gene Tsudik, Aur{\'e}lien Francillon, and Daniele Perito.
\newblock Smart: secure and minimal architecture for (establishing dynamic)
  root of trust.
\newblock In {\em Network and Distributed System Security (NDSS) Symposium},
  volume~12, pages 1--15, 2012.

\bibitem{fernandes2016security}
Earlence Fernandes, Jaeyeon Jung, and Atul Prakash.
\newblock Security analysis of emerging smart home applications.
\newblock In {\em 2016 IEEE Symposium on Security and Privacy (SP)}, pages
  636--654. IEEE, 2016.

\bibitem{funke2015end}
Sebastian Funke, J{\"o}rg Daubert, Alexander Wiesmaier, Panayotis Kikiras, and
  Max Muehlhaeuser.
\newblock End-2-end privacy architecture for iot.
\newblock In {\em 2015 IEEE Conference on Communications and Network Security
  (CNS)}, pages 705--706. IEEE, 2015.

\bibitem{garcia2016lock}
Flavio~D Garcia, David Oswald, Timo Kasper, and Pierre Pavlid{\`e}s.
\newblock Lock it and still lose it—on the (in) security of automotive remote
  keyless entry systems.
\newblock In {\em 25th USENIX Security Symposium (USENIX Sec)}, 2016.

\bibitem{gowsikhaa2012suspicious}
D~Gowsikhaa, S~Abirami, et~al.
\newblock {S}uspicious {H}uman {A}ctivity {D}etection {F}rom {S}urveillance
  {V}ideos.
\newblock {\em International Journal on Internet \& Distributed Computing
  Systems}, 2(2), 2012.

\bibitem{gross2017supervisory}
KC~Gross, K~Baclawski, ES~Chan, D~Gawlick, A~Ghoneimy, and ZH~Liu.
\newblock A supervisory control loop with prognostics for human-in-the-loop
  decision support and control applications.
\newblock In {\em 2017 IEEE conference on cognitive and computational aspects
  of situation management (CogSIMA)}, pages 1--7. IEEE, 2017.

\bibitem{grossmann2020cloudless}
Marcel Gro{\ss}mann and Christos Ioannidis.
\newblock Cloudless computing-a vision to become reality.
\newblock In {\em International Conference on Information Networking (ICOIN)},
  2020.

\bibitem{grossmann2019applicability}
Marcel Gro{\ss}mann, Christos Ioannidis, and Duy~Thanh Le.
\newblock Applicability of serverless computing in fog computing environments
  for iot scenarios.
\newblock In {\em IEEE/ACM International Conference on Utility and Cloud
  Computing Companion}, 2019.

\bibitem{guan2017trustshadow}
Le~Guan, Peng Liu, Xinyu Xing, Xinyang Ge, Shengzhi Zhang, Meng Yu, and Trent
  Jaeger.
\newblock {Trustshadow: Secure execution of unmodified applications with ARM
  TrustZone}.
\newblock In {\em Proceedings of the 15th Annual International Conference on
  Mobile Systems, Applications, and Services (Mobisys)}, pages 488--501, 2017.

\bibitem{guo2013opportunistic}
Bin Guo, Daqing Zhang, Zhu Wang, Zhiwen Yu, and Xingshe Zhou.
\newblock Opportunistic iot: Exploring the harmonious interaction between human
  and the internet of things.
\newblock In {\em Journal of Network and Computer Applications}, volume~36,
  pages 1531--1539. Elsevier, 2013.

\bibitem{habib2005photogrammetric}
Ayman Habib, Mwafag Ghanma, Michel Morgan, and Rami Al-Ruzouq.
\newblock Photogrammetric and lidar data registration using linear features.
\newblock {\em Photogrammetric Engineering \& Remote Sensing}, 71(6):699--707,
  2005.

\bibitem{hall1997introduction}
David~L Hall and James Llinas.
\newblock An introduction to multisensor data fusion.
\newblock {\em Proceedings of the IEEE}, 85(1):6--23, 1997.

\bibitem{hsiao2011h}
Yi-Mao Hsiao, Jeng-Farn Lee, Jai-Shiarng Chen, and Yuan-Sun Chu.
\newblock H. 264 video transmissions over wireless networks: Challenges and
  solutions.
\newblock {\em Computer Communications}, 34(14):1661--1672, 2011.

\bibitem{hu2018olympian}
Yitao Hu, Swati Rallapalli, Bongjun Ko, and Ramesh Govindan.
\newblock {O}lympian: {S}cheduling {G}pu {U}sage in a {D}eep {N}eural {N}etwork
  {M}odel {S}erving {S}ystem.
\newblock In {\em Proceedings of the 19th International Middleware Conference},
  pages 53--65. ACM, 2018.

\bibitem{hua2017vtz}
Zhichao Hua, Jinyu Gu, Yubin Xia, Haibo Chen, Binyu Zang, and Haibing Guan.
\newblock vtz: Virtualizing arm trustzone.
\newblock In {\em 26th USENIX Security Symposium (USENIX Sec)}, pages 541--556,
  2017.

\bibitem{hung2018videoedge}
Chien-Chun Hung, Ganesh Ananthanarayanan, Peter Bodik, Leana Golubchik, Minlan
  Yu, Paramvir Bahl, and Matthai Philipose.
\newblock Videoedge: Processing camera streams using hierarchical clusters.
\newblock In {\em 2018 IEEE/ACM Symposium on Edge Computing (SEC)}, pages
  115--131. IEEE, 2018.

\bibitem{ibrahim2017seed}
Ahmad Ibrahim, Ahmad-Reza Sadeghi, and Shaza Zeitouni.
\newblock Seed: secure non-interactive attestation for embedded devices.
\newblock In {\em Proceedings of the 10th ACM Conference on Security and
  Privacy in Wireless and Mobile Networks}, pages 64--74, 2017.

\bibitem{imteaj2019distributed}
Ahmed Imteaj and M~Hadi Amini.
\newblock Distributed sensing using smart end-user devices: pathway to
  federated learning for autonomous iot.
\newblock In {\em IEEE International Conference on Computational Science and
  Computational Intelligence (CSCI)}, 2019.

\bibitem{jayaraman2017privacy}
Prem~Prakash Jayaraman, Xuechao Yang, Ali Yavari, Dimitrios Georgakopoulos, and
  Xun Yi.
\newblock Privacy preserving internet of things: From privacy techniques to a
  blueprint architecture and efficient implementation.
\newblock {\em Future Generation Computer Systems}, 76:540--549, 2017.

\bibitem{jeffery2006declarative}
Shawn~R Jeffery, Gustavo Alonso, Michael~J Franklin, Wei Hong, and Jennifer
  Widom.
\newblock Declarative support for sensor data cleaning.
\newblock In {\em International Conference on Pervasive Computing}, pages
  83--100. Springer, 2006.

\bibitem{jiang2019differentially}
Linshan Jiang, Xin Lou, Rui Tan, and Jun Zhao.
\newblock Differentially private collaborative learning for the {IoT} edge.
\newblock In {\em EWSN}, 2019.

\bibitem{karimienergy}
Mohsen Karimi and Hyoseung Kim.
\newblock Energy scheduling for task execution on intermittently-powered
  devices.

\bibitem{kato2002obstacle}
Takeo Kato, Yoshiki Ninomiya, and Ichiro Masaki.
\newblock An obstacle detection method by fusion of radar and motion stereo.
\newblock {\em IEEE transactions on intelligent transportation systems},
  3(3):182--188, 2002.

\bibitem{kim2018securing}
Chung~Hwan Kim, Taegyu Kim, Hongjun Choi, Zhongshu Gu, Byoungyoung Lee, Xiangyu
  Zhang, and Dongyan Xu.
\newblock Securing real-time microcontroller systems through customized memory
  view switching.
\newblock In {\em Network and Distributed System Security (NDSS) Symposium},
  pages 1--15, 2018.

\bibitem{kolter2011redd}
J.~Kolter and M.~Johnson.
\newblock R{E}{D}{D}: A {P}ublic {D}ata {S}et for {E}nergy {D}isaggregation
  {R}esearch.
\newblock In {\em SustKDD}, 2011.

\bibitem{koo2017rl}
Jinkyu Koo, Xiaojun Lin, and Saurabh Bagchi.
\newblock {RL-BLH: Learning-based battery control for cost savings and privacy
  preservation for smart meters}.
\newblock In {\em 2017 47th Annual IEEE/IFIP International Conference on
  Dependable Systems and Networks (DSN)}, pages 519--530. IEEE, 2017.

\bibitem{Krebs2016DDoS}
Brian Krebs.
\newblock {DDoS on Dyn Impacts Twitter, Spotify, Reddit}.

\bibitem{lee2018pretzel}
Yunseong Lee, Alberto Scolari, Byung-Gon Chun, Marco~Domenico Santambrogio,
  Markus Weimer, and Matteo Interlandi.
\newblock {PRETZEL}: {O}pening the {B}lack {B}ox of {M}achine {L}earning
  {P}rediction {S}erving {S}ystems.
\newblock In {\em 13th USENIX Symposium on Operating Systems Design and
  Implementation (OSDI 18)}, pages 611--626, 2018.

\bibitem{li2019federated}
Tian Li, Anit~Kumar Sahu, Ameet Talwalkar, and Virginia Smith.
\newblock Federated learning: Challenges, methods, and future directions.
\newblock {\em arXiv preprint arXiv:1908.07873}, 2019.

\bibitem{liu2018epic}
Jianqing Liu, Chi Zhang, and Yuguang Fang.
\newblock Epic: A differential privacy framework to defend smart homes against
  internet traffic analysis.
\newblock {\em IEEE Internet of Things Journal}, 2018.

\bibitem{liu2016linear}
Shijie Liu, Xiaohua Tong, Jie Chen, Xiangfeng Liu, Wenzheng Sun, Huan Xie, Peng
  Chen, Yanmin Jin, and Zhen Ye.
\newblock A linear feature-based approach for the registration of unmanned
  aerial vehicle remotely-sensed images and airborne lidar data.
\newblock {\em Remote Sensing}, 8(2):82, 2016.

\bibitem{liu2018demand}
Sicong Liu, Yingyan Lin, Zimu Zhou, Kaiming Nan, Hui Liu, and Junzhao Du.
\newblock On-demand deep model compression for mobile devices: A usage-driven
  model selection framework.
\newblock In {\em Proceedings of the 16th Annual International Conference on
  Mobile Systems, Applications, and Services}, pages 389--400, 2018.

\bibitem{liu2019caesar}
Xiaochen Liu, Pradipta Ghosh, Oytun Ulutan, BS~Manjunath, Kevin Chan, and
  Ramesh Govindan.
\newblock Caesar: cross-camera complex activity recognition.
\newblock In {\em Proceedings of the 17th Conference on Embedded Networked
  Sensor Systems}, pages 232--244, 2019.

\bibitem{lu2019blockchain}
Yunlong Lu, Xiaohong Huang, Yueyue Dai, Sabita Maharjan, and Yan Zhang.
\newblock Blockchain and federated learning for privacy-preserved data sharing
  in industrial {IoT}.
\newblock {\em IEEE Trans. on Industrial Informatics}, 2019.

\bibitem{maeng2017alpaca}
Kiwan Maeng, Alexei Colin, and Brandon Lucia.
\newblock Alpaca: intermittent execution without checkpoints.
\newblock {\em Proceedings of the ACM on Programming Languages},
  1(OOPSLA):1--30, 2017.

\bibitem{maeng2018adaptive}
Kiwan Maeng and Brandon Lucia.
\newblock Adaptive dynamic checkpointing for safe efficient intermittent
  computing.
\newblock In {\em 13th USENIX Symposium on Operating Systems Design and
  Implementation (OSDI)}, pages 129--144, 2018.

\bibitem{maeng2020adaptive}
Kiwan Maeng and Brandon Lucia.
\newblock Adaptive low-overhead scheduling for periodic and reactive
  intermittent execution.
\newblock In {\em 41st ACM SIGPLAN Conference on Programming Language Design
  and Implementation (PLDI 2020)}, 2020.

\bibitem{car-hacking-2019}
Forbes Magazine.
\newblock Hacked driverless cars could cause collisions and gridlock in cities.
\newblock March 2019.

\bibitem{iot-video-attack-2018}
Wired Magazine.
\newblock An elaborate hack shows how much damage iot bugs can do.
\newblock April 2018.

\bibitem{HotelLockHack}
ZDNet Magazine.
\newblock Hackers built a 'master key' for millions of hotel rooms.
\newblock April 2018.

\bibitem{mclaughlin2011protecting}
S.~McLaughlin, P.~McDaniel, and W.~Aiello.
\newblock Protecting {C}onsumer {P}rivacy from {E}lectric {L}oad {M}onitoring.
\newblock In {\em CCS}, October 2011.

\bibitem{mettes2017spatial}
Pascal Mettes and Cees~GM Snoek.
\newblock {S}patial-aware {O}bject {E}mbeddings for {Z}ero-shot {L}ocalization
  and {C}lassification of {A}ctions.
\newblock In {\em Proceedings of the IEEE International Conference on Computer
  Vision}, pages 4443--4452, 2017.

\bibitem{mijumbi2016management}
Rashid Mijumbi, Joan Serrat, Juan-Luis Gorricho, Steven Latr{\'e}, Marinos
  Charalambides, and Diego Lopez.
\newblock Management and orchestration challenges in network functions
  virtualization.
\newblock {\em IEEE Communications Magazine}, 54(1):98--105, 2016.

\bibitem{molina2010private}
A.~Molina-Markham, P.~Shenoy, K.~Fu, E.~Cecchet, and D.~Irwin.
\newblock Private {Me}moirs of a {S}mart {M}eter.
\newblock In {\em BuildSys}, 2010.

\bibitem{muench2018you}
Marius Muench, Jan Stijohann, Frank Kargl, Aur{\'e}lien Francillon, and Davide
  Balzarotti.
\newblock What you corrupt is not what you crash: Challenges in fuzzing
  embedded devices.
\newblock In {\em Network and Distributed System Security (NDSS) Symposium},
  pages 1--15, 2018.

\bibitem{neumayer2010network}
Sebastian Neumayer and Eytan Modiano.
\newblock Network reliability with geographically correlated failures.
\newblock In {\em 2010 Proceedings IEEE INFOCOM}, pages 1--9. IEEE, 2010.

\bibitem{nithin2017globality}
Kanishka Nithin and Fran{\c{c}}ois Br{\'e}mond.
\newblock {{G}}lobality--locality-based {{C}}onsistent {{D}}iscriminant
  {{F}}eature {{E}}nsemble for {{M}}ulticamera {{T}}racking.
\newblock {\em IEEE Transactions on Circuits and Systems for Video Technology},
  27(3):431--440, 2017.

\bibitem{nunes2015survey}
David~Sousa Nunes, Pei Zhang, and Jorge~S{\'a} Silva.
\newblock A survey on human-in-the-loop applications towards an internet of
  all.
\newblock {\em IEEE Communications Surveys \& Tutorials}, 17(2):944--965, 2015.

\bibitem{nunes2019vrased}
Ivan De~Oliveira Nunes, Karim Eldefrawy, Norrathep Rattanavipanon, Michael
  Steiner, and Gene Tsudik.
\newblock $\{$VRASED$\}$: A verified hardware/software co-design for remote
  attestation.
\newblock In {\em 28th USENIX Security Symposium}, pages 1429--1446, 2019.

\bibitem{ojanpera2018use}
Tiia Ojanper{\"a}, Jukka M{\"a}kel{\"a}, Olli M{\"a}mmel{\"a}, Mikko Majanen,
  and Ossi Martikainen.
\newblock Use cases and communications architecture for 5g-enabled road safety
  services.
\newblock In {\em 2018 European Conference on Networks and Communications
  (EuCNC)}, pages 335--340. IEEE, 2018.

\bibitem{pakha2018reinventing}
Chrisma Pakha, Aakanksha Chowdhery, and Junchen Jiang.
\newblock Reinventing video streaming for distributed vision analytics.
\newblock In {\em 10th $\{$USENIX$\}$ Workshop on Hot Topics in Cloud Computing
  (HotCloud 18)}, 2018.

\bibitem{panta2011efficient}
Rajesh~Krishna Panta, Saurabh Bagchi, and Samuel~P Midkiff.
\newblock Efficient incremental code update for sensor networks.
\newblock {\em ACM Transactions on Sensor Networks (TOSN)}, 7(4):1--32, 2011.

\bibitem{parmar2012adobe}
H~Parmar and M~Thornburgh.
\newblock Adobe’s real time messaging protocol.
\newblock {\em Copyright Adobe Systems Incorporated}, pages 1--52, 2012.

\bibitem{pyrgelis2017knock}
Apostolos Pyrgelis, Carmela Troncoso, and Emiliano De~Cristofaro.
\newblock Knock knock, who's there? membership inference on aggregate location
  data.
\newblock 2018.

\bibitem{qi2016quantifying}
Yinan Qi, Mythri Hunukumbure, Maziar Nekovee, Javier Lorca, and Victoria
  Sgardoni.
\newblock Quantifying data rate and bandwidth requirements for immersive 5g
  experience.
\newblock In {\em 2016 IEEE International Conference on Communications
  Workshops (ICC)}, pages 455--461. IEEE, 2016.

\bibitem{Qiu18d}
Hang Qiu, Fawad Ahmad, Fan Bai, Marco Gruteser, and Ramesh Govindan.
\newblock Avr: Augmented vehicular reality.
\newblock In {\em Proceedings of the 16th Annual International Conference on
  Mobile Systems, Applications, and Services (Mobisys)}, pages 81--95, 2018.

\bibitem{ragland2014survey}
Kirubaraj Ragland and P~Tharcis.
\newblock A survey on object detection, classification and tracking methods.
\newblock {\em Int. J. Eng. Res. Technol}, pages 622--628, 2014.

\bibitem{ristani2018features}
Ergys Ristani and Carlo Tomasi.
\newblock {F}eatures for {M}ulti-target {M}ulti-camera {T}racking and
  {R}e-identification.
\newblock {\em arXiv preprint arXiv:1803.10859}, 2018.

\bibitem{freertos-mpu}
Free RTOS.
\newblock {The FreeRTOS Kernel MPU Support}.

\bibitem{sattler2019robust}
Felix Sattler, Simon Wiedemann, Klaus-Robert M{\"u}ller, and Wojciech Samek.
\newblock Robust and communication-efficient federated learning from non-{IID}
  data.
\newblock {\em IEEE transactions on neural networks and learning systems},
  2019.

\bibitem{shacham2004effectiveness}
Hovav Shacham, Matthew Page, Ben Pfaff, Eu-Jin Goh, Nagendra Modadugu, and Dan
  Boneh.
\newblock On the effectiveness of address-space randomization.
\newblock In {\em Proceedings of the 11th ACM conference on Computer and
  communications security}, pages 298--307, 2004.

\bibitem{sharma2010sensor}
Abhishek~B Sharma, Leana Golubchik, and Ramesh Govindan.
\newblock Sensor faults: Detection methods and prevalence in real-world
  datasets.
\newblock {\em ACM Transactions on Sensor Networks (TOSN)}, 6(3):1--39, 2010.

\bibitem{shi2016edge}
Weisong Shi, Jie Cao, Quan Zhang, Youhuizi Li, and Lanyu Xu.
\newblock Edge computing: Vision and challenges.
\newblock {\em IEEE internet of things journal}, 3(5):637--646, 2016.

\bibitem{shou2018online}
Zheng Shou, Junting Pan, Jonathan Chan, Kazuyuki Miyazawa, Hassan Mansour,
  Anthony Vetro, Xavier Giro-i Nieto, and Shih-Fu Chang.
\newblock {O}nline {D}etection of {A}ction {S}tart in {U}ntrimmed, {S}treaming
  {V}ideos.
\newblock In {\em Proceedings of the European Conference on Computer Vision
  (ECCV)}, pages 534--551, 2018.

\bibitem{solera2016tracking}
Francesco Solera, Simone Calderara, Ergys Ristani, Carlo Tomasi, and Rita
  Cucchiara.
\newblock {{T}}racking {{S}}ocial {{G}}roups {{W}}ithin and {{A}}cross
  {{C}}ameras.
\newblock {\em IEEE Transactions on Circuits and Systems for Video Technology},
  2016.

\bibitem{sommer2013minerva}
Philipp Sommer and Branislav Kusy.
\newblock Minerva: Distributed tracing and debugging in wireless sensor
  networks.
\newblock In {\em the 11th ACM Conf. on Embedded Networked Sensor Systems
  (Sensys)}, pages 1--14, 2013.

\bibitem{song2016robust}
Haryong Song, Wonsub Choi, and Haedong Kim.
\newblock Robust vision-based relative-localization approach using an rgb-depth
  camera and lidar sensor fusion.
\newblock {\em IEEE Transactions on Industrial Electronics}, 63(6):3725--3736,
  2016.

\bibitem{sundaram2011demo}
Vinaitheerthan Sundaram, Patrick Eugster, and Xiangyu Zhang.
\newblock Demo abstract: Diagnostic tracing of wireless sensor networks with
  tinytracer.
\newblock In {\em 10th ACM/IEEE International Conference on Information
  Processing in Sensor Networks}, pages 145--146. IEEE, 2011.

\bibitem{sundaram2012prius}
Vinaitheerthan Sundaram, Patrick Eugster, and Xiangyu Zhang.
\newblock Prius: Generic hybrid trace compression for wireless sensor networks.
\newblock In {\em Proceedings of the 10th ACM Conference on Embedded Network
  Sensor Systems}, pages 183--196, 2012.

\bibitem{szewczyk2004lessons}
Robert Szewczyk, Joseph Polastre, Alan Mainwaring, and David Culler.
\newblock Lessons from a sensor network expedition.
\newblock In {\em European Workshop on Wireless Sensor Networks}, pages
  307--322. Springer, 2004.

\bibitem{tancreti2011aveksha}
Matthew Tancreti, Mohammad~Sajjad Hossain, Saurabh Bagchi, and Vijay
  Raghunathan.
\newblock Aveksha: A hardware-software approach for non-intrusive tracing and
  profiling of wireless embedded systems.
\newblock In {\em Proc. of the 9th ACM Conference on Embedded Networked Sensor
  Systems (Sensys)}, pages 288--301, 2011.

\bibitem{tancreti2015tardis}
Matthew Tancreti, Vinaitheerthan Sundaram, Saurabh Bagchi, and Patrick Eugster.
\newblock Tardis: software-only system-level record and replay in wireless
  sensor networks.
\newblock In {\em Proc. of the 14th Intl. Conf. on Information Processing in
  Sensor Networks (IPSN)}, pages 286--297, 2015.

\bibitem{tayyaba2017software}
Sahrish~Khan Tayyaba, Munam~Ali Shah, Omair~Ahmad Khan, and Abdul~Wahab Ahmed.
\newblock Software defined network (sdn) based internet of things (iot) a road
  ahead.
\newblock In {\em the International Conference on Future Networks and
  Distributed Systems}, pages 1--8, 2017.

\bibitem{tesfaye2017multi}
Yonatan~Tariku Tesfaye, Eyasu Zemene, Andrea Prati, Marcello Pelillo, and
  Mubarak Shah.
\newblock {{M}}ulti-target {{T}}racking in {{M}}ultiple {{N}}on-overlapping
  {{C}}ameras using {{C}}onstrained {{D}}ominant {{S}}ets.
\newblock {\em arXiv preprint arXiv:1706.06196}, 2017.

\bibitem{tian2017smartauth}
Yuan Tian, Nan Zhang, Yueh-Hsun Lin, XiaoFeng Wang, Blase Ur, Xianzheng Guo,
  and Patrick Tague.
\newblock Smartauth: User-centered authorization for the internet of things.
\newblock In {\em 26th $\{$USENIX$\}$ Security Symposium ($\{$USENIX$\}$
  Security 17)}, pages 361--378, 2017.

\bibitem{truong2016reputation}
Nguyen~B Truong, Tai-Won Um, and Gyu~Myoung Lee.
\newblock A reputation and knowledge based trust service platform for
  trustworthy social internet of things.
\newblock {\em Innovations in clouds, Internet \& networks (ICIN)}, 2016.

\bibitem{ulutan2018actor}
Oytun Ulutan, Swati Rallapalli, Carlos Torres, Mudhakar Srivatsa, and
  BS~Manjunath.
\newblock {A}ctor {C}onditioned {A}ttention {M}aps for {V}ideo {A}ction
  {D}etection.
\newblock In {\em the IEEE Winter Conference on Applications of Computer Vision
  (WACV)}. IEEE, 2020.

\bibitem{van2016intermittent}
Joel Van Der~Woude and Matthew Hicks.
\newblock Intermittent computation without hardware support or programmer
  intervention.
\newblock In {\em 12th USENIX Symposium on Operating Systems Design and
  Implementation (OSDI)}, pages 17--32, 2016.

\bibitem{wang2018bandwidth}
Junjue Wang, Ziqiang Feng, Zhuo Chen, Shilpa George, Mihir Bala, Padmanabhan
  Pillai, Shao-Wen Yang, and Mahadev Satyanarayanan.
\newblock Bandwidth-efficient live video analytics for drones via edge
  computing.
\newblock In {\em IEEE/ACM Symp. on Edge Computing (SEC)}, pages 159--173.
  IEEE, 2018.

\bibitem{wang2006reprogramming}
Qiang Wang, Yaoyao Zhu, and Liang Cheng.
\newblock Reprogramming wireless sensor networks: challenges and approaches.
\newblock {\em IEEE network}, 20(3):48--55, 2006.

\bibitem{wark2007real}
Tim Wark, Peter Corke, Johannes Karlsson, Pavan Sikka, and Philip Valencia.
\newblock Real-time image streaming over a low-bandwidth wireless camera
  network.
\newblock In {\em 3rd Intl. Conf. on Intelligent Sensors, Sensor Networks and
  Information}, pages 113--118, 2007.

\bibitem{xiong2018enhancing}
Jinbo Xiong, Jun Ren, Lei Chen, Zhiqiang Yao, Mingwei Lin, Dapeng Wu, and Ben
  Niu.
\newblock Enhancing privacy and availability for data clustering in intelligent
  electrical service of {IoT}.
\newblock {\em IEEE IoT Journal}, 2018.

\bibitem{xu2018distilling}
Chugui Xu, Ju~Ren, Deyu Zhang, and Yaoxue Zhang.
\newblock Distilling at the edge: A local differential privacy obfuscation
  framework for {IoT} data analytics.
\newblock {\em IEEE Communications Magazine}, 2018.

\bibitem{xu2017cross}
Yuanlu Xu, Xiaobai Liu, Lei Qin, and Song-Chun Zhu.
\newblock {{C}}ross-view {{P}}eople {{T}}racking by {{S}}cene-centered
  {{S}}patio-temporal {{P}}arsing.
\newblock In {\em AAAI}, pages 4299--4305, 2017.

\bibitem{yang2012minimizing}
Weining Yang, Ninghui Li, Yuan Qi, Wahbeh Qardaji, Stephen McLaughlin, and
  Patrick McDaniel.
\newblock Minimizing private data disclosures in the smart grid.
\newblock In {\em ACM conference on Computer and Communications Security},
  2012.

\bibitem{yao2019eugene}
Shuochao Yao, Yifan Hao, Yiran Zhao, Ailing Piao, Huajie Shao, Dongxin Liu,
  Shengzhong Liu, Shaohan Hu, Dulanga Weerakoon, Kasthuri Jayarajah, et~al.
\newblock Eugene: Towards deep intelligence as a service.
\newblock In {\em 2019 IEEE 39th International Conference on Distributed
  Computing Systems (ICDCS)}, pages 1630--1640. IEEE, 2019.

\bibitem{yao2019stardust}
Shuochao Yao, Tianshi Wang, Jinyang Li, and Tarek Abdelzaher.
\newblock Stardust: A deep learning serving system in iot: demo abstract.
\newblock In {\em Proceedings of the 17th Conference on Embedded Networked
  Sensor Systems}, pages 402--403, 2019.

\bibitem{yao2018fastdeepiot}
Shuochao Yao, Yiran Zhao, Huajie Shao, ShengZhong Liu, Dongxin Liu, Lu~Su, and
  Tarek Abdelzaher.
\newblock Fastdeepiot: Towards understanding and optimizing neural network
  execution time on mobile and embedded devices.
\newblock In {\em Proceedings of the 16th ACM Conference on Embedded Networked
  Sensor Systems}, pages 278--291, 2018.

\bibitem{yao2017deepiot}
Shuochao Yao, Yiran Zhao, Aston Zhang, Lu~Su, and Tarek Abdelzaher.
\newblock Deepiot: Compressing deep neural network structures for sensing
  systems with a compressor-critic framework.
\newblock In {\em Proceedings of the 15th ACM Conference on Embedded Network
  Sensor Systems}, pages 1--14, 2017.

\bibitem{yick2008wireless}
Jennifer Yick, Biswanath Mukherjee, and Dipak Ghosal.
\newblock Wireless sensor network survey.
\newblock {\em Computer networks}, 52(12):2292--2330, 2008.

\bibitem{yu2020learning}
Tianlong Yu, Tian Li, Yuqiong Sun, Susanta Nanda, Virginia Smith, Vyas Sekar,
  and Srinivasan Seshan.
\newblock Learning context-aware policies from multiple smart homes via
  federated multi-task learning.
\newblock In {\em IEEE/ACM International Conference on Internet of Things
  Design and Implementation (IoTDI)}, 2020.

\bibitem{yue2019data}
Rui Yue, Hao Xu, Jianqing Wu, Renjuan Sun, and Changwei Yuan.
\newblock Data registration with ground points for roadside lidar sensors.
\newblock {\em Remote Sensing}, 11(11):1354, 2019.

\bibitem{zaddach2014avatar}
Jonas Zaddach, Luca Bruno, Aurelien Francillon, Davide Balzarotti, et~al.
\newblock Avatar: A framework to support dynamic security analysis of embedded
  systems' firmwares.
\newblock In {\em Network and Distributed System Security (NDSS) Symposium},
  volume~14, pages 1--16, 2014.

\bibitem{zhang2017live}
Haoyu Zhang, Ganesh Ananthanarayanan, Peter Bodik, Matthai Philipose, Paramvir
  Bahl, and Michael~J Freedman.
\newblock Live video analytics at scale with approximation and delay-tolerance.
\newblock In {\em 14th $\{$USENIX$\}$ Symposium on Networked Systems Design and
  Implementation ($\{$NSDI$\}$ 17)}, pages 377--392, 2017.

\bibitem{zhao2019exploring}
Junxuan Zhao.
\newblock {\em Exploring the fundamentals of using infrastructure-based LiDAR
  sensors to develop connected intersections}.
\newblock PhD thesis, 2019.

\bibitem{zhao2019detection}
Junxuan Zhao, Hao Xu, Hongchao Liu, Jianqing Wu, Yichen Zheng, and Dayong Wu.
\newblock Detection and tracking of pedestrians and vehicles using roadside
  lidar sensors.
\newblock {\em Transportation research part C: emerging technologies},
  100:68--87, 2019.

\bibitem{zhao2019mobile}
Yang Zhao, Jun Zhao, Linshan Jiang, Rui Tan, and Dusit Niyato.
\newblock Mobile edge computing, blockchain and reputation-based crowdsourcing
  {IOT} federated learning: A secure, decentralized and privacy-preserving
  system.
\newblock {\em arXiv preprint arXiv:1906.10893}, 2019.

\bibitem{zhao2017temporal}
Yue Zhao, Yuanjun Xiong, Limin Wang, Zhirong Wu, Xiaoou Tang, and Dahua Lin.
\newblock {T}emporal {A}ction {D}etection with {S}tructured {S}egment
  {N}etworks.
\newblock In {\em Proceedings of the IEEE International Conference on Computer
  Vision}, pages 2914--2923, 2017.

\bibitem{zheng2018user}
Serena Zheng, Noah Apthorpe, Marshini Chetty, and Nick Feamster.
\newblock User perceptions of smart home iot privacy.
\newblock {\em Proceedings of the ACM on Human-Computer Interaction},
  2(CSCW):1--20, 2018.

\end{thebibliography}
}

\end{document}